\documentclass[pra,twocolumn,amsmath,amssymb,showpacs,superscriptaddress]{revtex4}
\usepackage{graphicx,epstopdf,bm}

\newcommand{\Fkt}[1]{\,\mathsf {#1}}
\ifx\Tr\renewcommand{\Tr}{\Fkt{Tr}}
\else\newcommand{\Tr}{\Fkt{Tr}}
\fi
\newcommand{\bra}[1]{\ensuremath{\langle#1|}}
\newcommand{\ket}[1]{\ensuremath{|#1\rangle}}

\begin{document}

\title{Efficient Monte Carlo characterization of quantum operations
  for qudits}

\author{Giulia Gualdi}
\affiliation{Dipartimento di Fisica ed Astronomia, Universit\`a di
  Firenze, Via Sansone 1, 50019 Sesto Fiorentino, Italy}
\affiliation{QSTAR, Largo Enrico Fermi 2, 50125 Firenze, Italy}

\author{David Licht}
\affiliation{Theoretische Physik, Universit\"{a}t Kassel,
  Heinrich-Plett-Str. 40, D-34132 Kassel, Germany} 

\author{Daniel M. Reich}
\affiliation{Theoretische Physik, Universit\"{a}t Kassel,
  Heinrich-Plett-Str. 40, D-34132 Kassel, Germany} 

\author{Christiane P. Koch}
\affiliation{Theoretische Physik, Universit\"{a}t Kassel,
  Heinrich-Plett-Str. 40, D-34132 Kassel, Germany} 
\email{christiane.koch@uni-kassel.de}

\begin{abstract}
  For qubits, Monte Carlo estimation of the average fidelity of
  Clifford unitaries is efficient -- it requires a number of
  experiments that is independent of the number  $n$ of qubits and
  classical computational 
  resources that scale only polynomially in $n$.  
  Here, we identify the requirements for efficient Monte
  Carlo estimation and the corresponding properties of the measurement
  operator basis when replacing two-level qubits by $p$-level
  qudits. Our analysis illuminates the intimate connection between
  mutually unbiased measurements and the existence of unitaries that
  can be characterized efficiently. It allows us to propose a
  'hierarchy' of generalizations of the standard Pauli basis from
  qubits to qudits according to the associated scaling of resources
  required in Monte Carlo estimation of the average fidelity. 
\end{abstract}


\date{\today}
\maketitle

\section{Introduction}
\label{sec:intro}

The capability to verify that a quantum operation has been properly
implemented is an important building block for quantum
technologies~\cite{NielsenChuang}. It requires evaluation of 
 suitable performance measures such as
the average fidelity or the worst case fidelity. In general, evaluating either 
measure scales very
unfavorably in system size due to the exponential scaling of the Hilbert
space dimension $d$ with the number $n$ of information carriers. 
Stochastic sampling techniques have recently allowed for 
impressive progress at reducing the resources required for determining
the average gate fidelity for qubits~\cite{FlammiaPRL11,daSilvaPRL11,ShabaniPRL11,SchmiegelowPRL11,MagesanPRL11,MagesanPRA12}.  
For example, Monte Carlo estimation can be employed to determine 
the average $n$-qubit gate fidelity $F_{av}$~\cite{FlammiaPRL11,daSilvaPRL11}. 
To this end, $F_{av}$ is expressed either 
in terms of the entanglement fidelity~\cite{daSilvaPRL11, FlammiaPRL11}
or as a sum over $d(d+1)$ state fidelities in
$d$-dimensional Hilbert space where the $d(d+1)$ states form a so-called
state 2-design~\cite{BenderskyPRL08,ReichPRL13}. The latter represents the
optimal strategy in terms of the average number of experiments that
need to be performed, the number of settings from which an experiment
is drawn in the Monte Carlo procedure and the associated computational
complexity~\cite{ReichPRL13}. The effort for estimating the average
gate fidelity can be further reduced when determining bounds instead
of $F_{av}$ itself~\cite{ReichPRL13}. The bounds are given by two
classical fidelities in Hilbert space each made up of $d$ state fidelities~\cite{HofmannPRL05}.

These statements hold for both general unitaries and Clifford
gates. However, for Clifford gates, the three approaches differ merely
in the number of experimental settings; the average number of
experiments is independent of system
size~\cite{FlammiaPRL11,daSilvaPRL11,ReichPRL13}. As a consequence,
estimating the average fidelity of a Clifford gate is a task that can
be performed efficiently, i.e., with an effort that scales at most
polynomially with the number of  qubits. 

Clifford gates represent an important subset
of quantum gates -- they facilitate fault-tolerant
computation~\cite{GottesmanChaos99} and yield a universal set when 
augmented by the proper local phasegate~\cite{Boykin00}. They
can be used to prepare entangled states and perform quantum
teleportation even though their computing power is not stronger than
classical~\cite{GottesmanKnill}. 
The striking observation that the experimental effort for Clifford gate
characterization does not scale exponentially with the number of qubits 
is due to the property of Clifford gates to map stabilizer states
into stabilizer states. This property is also exploited by
another efficient method for determining the average gate fidelity,
termed randomized benchmarking~\cite{MagesanPRL11,MagesanPRA12}. 

The Clifford gate property translates, for Monte Carlo estimation of
the average fidelity, into a relevance
distribution which is uniform and known \textit{a
  priori}~\cite{FlammiaPRL11,daSilvaPRL11}. A uniform relevance
distribution does not require 
sampling; and the average number of experiments becomes
independent of system size. It turns out, however, that the uniformity
of the relevance distribution is tied to the Pauli operators having
eigenvalues $\pm 1$.
It therefore applies to qubits but not  to Hilbert spaces of 
prime power dimensions $d=p^n$ with $p$ other than two. 
This raises
the question of whether and how the Clifford property of mapping
stabilizer states into stabilizer states can be exploited to
efficiently estimate the average gate fidelity for qudits ($p>2$). 

Qudits in general and qutrits ($p=3$) in particular occur naturally in
many quantum systems: They can be encoded in anharmonic ladders of
e.g. superconducting circuits~\cite{NeeleySci09,StrauchPRL12},  
in orbital angular momentum modes of
photons~\cite{MolinaPRL04,GroeblacherNJP06} 
or in the polarization of biphotons~\cite{RalphPRA07,LanyonPRL08}. 
Compared to qubits as quantum information carriers, they offer
advantages  in terms of
increased security and higher channel capacity in quantum
communication and better efficiency in quantum
information, see
e.g. Refs.~\cite{MolinaPRL04,GroeblacherNJP06,RalphPRA07}.  Since
device characterization is one of the prerequisites for any
quantum information and communication architecture, 
it would represent a severe disadvantage of qudits if the average
fidelity of qudit Clifford gates could not be determined efficiently. 

Here, we demonstrate that Monte Carlo estimation of the average fidelity can
be made efficient for Clifford gates of qudits by suitable choice of
the operator basis for the measurements. 
Based on intuition obtained for the qubit case, we show that the
measurement basis needs to allow for a partitioning into $d+1$
commuting sets of operators to ensure existence of a non-trivial class
of unitaries that map stabilizer states into stabilizer states and
yield a uniform relevance distribution. For qudits ($p>2$), only
unitary, non-Hermitian operators give rise to such a maximal
partitioning. Two routes can be followed to obtain a practical
characterization protocol from this observation: One can either 
construct Hermitian operators by suitable superposition of the basis
unitaries; 
or utilize the concept of quantum circuits to simulate Hermitian
measurements. We discuss both options. In general, we show that one
can define a hierarchy of operator bases  according to their scaling
of resources  in the Monte Carlo characterization of Clifford gates.

The paper is organized as follows: We start with a review of
Monte Carlo estimation of the average fidelity for
qubits~\cite{FlammiaPRL11,daSilvaPRL11}  in  
Section~\ref{sec:MC}. In particular, we explain the role of operator
bases of Hilbert space for evaluating the relevance distribution for
qubits and we show how the scaling in resources is obtained from
it. We construct the operator basis for qudits in Sec.~\ref{sec:opbases},
starting from the condition of a maximal partitioning and imposing
further constraints on the operators to ensure efficient
characterization for a maximal number of unitaries. 
We present the relevance distributions
resulting from these bases and discuss the corresponding Monte Carlo
procedures in Sec.~\ref{sec:qudit_reldist}. Section~\ref{sec:concl}
concludes. 

\section{Monte Carlo estimation of the average fidelity 
  for qubit Clifford gates} 
\label{sec:MC}

We first provide an overview over the general ideas underlying
the Monte Carlo
approach~\cite{FlammiaPRL11,daSilvaPRL11}. Subsequently we 
explain, following Ref.~\cite{FlammiaPRL11}, why for a Clifford gate 
the resources required for Monte Carlo estimation of the average
fidelity do not  scale exponentially with the number of qubits.

\subsection{Recasting $F_{av}$ in terms of measurements}
\label{subsec:MC}

We consider a system of $n$ qubits with a Hilbert space of dimension
$d=2^n$. The associated Liouville space, of dimension $d^2$, can be
spanned by a complete and orthonormal operator basis
$W_k$ with $\Tr[W_iW_k]=d\delta_{i,k}$ $\forall i,k=1,\ldots,d^2$. 
From a physical perspective,
the operator basis represents the set of measurements that will have
to be performed. The goal is  to estimate the average fidelity
$F_{av}$ of a quantum device that is supposed to execute the gate
$U\in\mathcal{U}(d)$. In other words, determining $F_{av}$ verifies
how well the actual evolution of the system, 
represented by the dynamical map $\mathcal{D}$, matches 
the target $U$~\cite{NielsenChuang}.

One possibility to evaluate $F_{av}$ with a Monte Carlo
procedure~\cite{FlammiaPRL11} rewrites $F_{av}$ in terms of the entanglement
fidelity $F_e$~\cite{HorodeckiPRA99,NielsenPLA02},
\begin{equation}\label{eq:favfe}
F_{av}=\frac{dF_e+1}{d+1}\,.
\end{equation}
$F_e$ is defined as~\cite{NielsenPLA02,SchumacherPRA96,FlammiaPRL11}
\begin{equation}\label{eq:ffe}
   F_{e}=\frac{1}{d^2}\Tr\left[\mathcal{U}^\dagger\mathcal{D}\right]\,,
\end{equation}
where $\mathcal{U}$ denotes the unitary dynamical map corresponding to
the desired gate $U$. A second option, using the channel-state
isomorphism,  interprets $F_e$ as a state fidelity
on an extended $d^4$-dimensional Liouville space~\cite{daSilvaPRL11}.
The two approaches are equivalent.
Expanding the trace in Eq.~\eqref{eq:ffe} in the  
operator basis $W_k$, one obtains~\cite{FlammiaPRL11}
\begin{equation}
\label{eq:feop}
F_e=\frac{1}{d^4}\sum_{k,k'}\Tr
[W_kUW_{k'}U^\dagger]\Tr[W_k\mathcal{D}(W_{k'})]\,. 
\end{equation}
The corresponding measurements are performed on inputs 
that have passed the device. Both are subjected to Monte Carlo
sampling. Formally,  the inputs are the operators
$W_{k'}$.  The obstacle that, in an 
experiment, one cannot  prepare  input operators is circumvented by
sampling, additionally, over each input operator's
eigenstates~\cite{FlammiaPRL11}.  The set of
inputs $I$ consists of all $T=d^2$ (rescaled) operators $W_{k'}/d$ that 
constitute the orthonormal basis. In practical terms, Monte Carlo
estimation of the average fidelity 
consists in randomly selecting pairs of input states and measurements
that will be performed on the output obtained after sending the input
through the quantum device. Summing up all measurement outcomes with
the appropriate weights, given by the so-called relevance 
distribution (for details see Sec.~\ref{subsec:Pr} below), yields 
the average fidelity.

The formal use of input operators, or, equivalently, the
channel-state isomorphism, can be avoided by evaluating 
$F_{av}$ as a state 2-design~\cite{BenderskyPRL08,ReichPRL13}. Then
the set of inputs $I$ consists of $T=d(d+1)$ regular Hilbert space
states, which make up $d+1$ mutually unbiased bases (MUB), and the
average fidelity is expressed as 
\begin{eqnarray}
  \label{eq:F_av_2design}
  F_{av}&=&\frac{1}{d(d+1)}\sum_{j=1}^{d(d+1)}
  \Tr\left[\rho_j^{ideal}\rho_j^{actual}\right] \\
  &=& \nonumber
  \frac{1}{d^2(d+1)}\sum_{j=1}^{d(d+1)}\sum_{k=1}^{d^2}
  \Tr\left[\rho_j^{ideal}W_k\right]\Tr\left[\rho_j^{actual}W_k\right] 
  \,,
\end{eqnarray}
where $\rho_j^{ideal}=U\ket{\Psi_j}\bra{\Psi_j}U^+$ 
and $\rho_j^{actual}=\mathcal{D}(\ket{\Psi_j}\bra{\Psi_j})$.
Another option is to determine bounds on the average gate
fidelity instead of $F_{av}$ itself using two classical
fidelities~\cite{HofmannPRL05,ReichPRL13}. 
Each classical fidelity is expressed as a sum over $T=d$ input
states, analogously to Eq.~\eqref{eq:F_av_2design}, with the states
belonging to two MUB~\cite{ReichPRL13}.
The different sets of inputs for the three protocols result 
in different numbers of required experimental settings, average
numbers of actual measurements, and classical computational
resources~\cite{ReichPRL13}. 

\subsection{Relevance distribution}
\label{subsec:Pr}

The idea underlying the Monte Carlo approach is to treat 
$\Tr\left[W_k\mathcal{D}(I_i)\right]$, where $I_i\in I$ denotes an
element of the set of inputs, either operators or states,
as a random variable. Then the
average fidelity becomes the expectation value of a random
variable, i.e., one can write $F_{av}$ as 
\begin{equation}\label{eq:fav}
  F_{av}^j=\sum_{i=1}^T\sum_{k=1}^{d^2}P^{j}(i,k)X_{i,k} \,,
\end{equation}
where $j$ indicates the specific protocol (entanglement fidelity, 
state 2-design, or classical fidelities). 
$P^{j}(i,k)$ is the so-called relevance (i.e.,
probability) distribution associated to  $F_{av}^j$, and the 
$X_{i,k}$ are the values taken by the random variable $X$. Obviously,
$\Tr\left[W_kUI_iU^\dagger\right]$ will be absorbed into
$P^{j}(i,k)$. The indices $i\in[1,T]$ and $k\in[1,d^2]$ 
run over the set of inputs and the set of measurements.
The size of the space of Monte Carlo events, i.e., the
domain of the relevance distribution, is therefore given by $Td^2$.
The relevance distribution $P^{j}(i,k)$
and random variable $X_{i,k}$ can be expressed 
in terms of the characteristic functions,  
\begin{subequations}\label{eq:chi}
\begin{eqnarray}\label{eq:chi_U}
\chi^{j}_{U}(i,k) &=&\Tr\left[W_kUI_iU^\dagger\right]\,,\\  
\chi^{j}_{\mathcal{D}}(i,k) &=&
\Tr\left[W_k\mathcal{D}(I_i)\right]\,, \label{eq:chi_D}
\end{eqnarray}  
\end{subequations}
that represent the expectation value
of the $k$th measurement after the $i$th
input has passed the device. This allows to write
\begin{subequations}\label{eq:Prel_Xik}
\begin{eqnarray}
\label{eq:Xik}
X_{ik} &=& \frac{\chi_\mathcal{D}^{j}(i,k)}{\chi_U^{j}(i,k)}\,,  \\
\label{eq:rel}
P^{j}(i,k)&=&\frac{1}{\mathcal{N}} \left[\chi_U^{j}(i,k)\right]^2
\end{eqnarray}  
\end{subequations}
with $\mathcal{N}$ ensuring proper normalization:
$\mathcal{N}=d^2$
for the protocols based on the entanglement fidelity and on two
classical fidelities, whereas  $\mathcal{N}=d^2(d+1)$ for the protocol
employing a state 2-design. 

When evaluating  $F_{av}^j$ as
expectation value of the random variable $X$ taking values
$X_{i,k}$ with known probability $P^{j}(i,k)$, one is
faced with the problem that the $X_{i,k}$ cannot be accessed
directly. As can be seen from Eq.~\eqref{eq:Xik}, 
they depend on another random variable, the expectation value 
$\Tr\left[W_k\mathcal{D}(I_i)\right]$ of $W_k$. Due to the statistical
nature of quantum measurements as well as 
random errors in the experiment, it will be necessary to make repeated
measurements to determine $X_{i,k}$. We assume for a moment 
that the $X_{i,k}$ have been determined with
sufficient accuracy and explain below what this assumption entails.
Provided the $X_{i,k}$ are known, Monte Carlo sampling estimates the
expectation value $F_{av}^j$ of the random variable $X$ by a finite
number of realizations, 
\begin{equation}
  \label{eq:MCapprox}
  F_{av}^j=\lim_{L\to\infty} F_{L} \quad\mathrm{with}\quad
  F_{L} = \frac{1}{L}\sum_{l=1}^L X_{\kappa_l}\,.
\end{equation}
Here, $\kappa_l$ is the index corresponding
to the $l$th input-output pair, i.e., $\kappa_l=(i_l,k_l)$. It 
can take on $Td^2$ values. The sample size $L$ is chosen to 
guarantee that the probability for $F_{L}$ to differ from 
$F_{av}^j$ by more than $\epsilon$ is less than $\delta$. 
The key point of the Monte Carlo approach is that while the size of
the event space scales with the system size $d$, 
$L$ depends only on the desired accuracy 
$\epsilon$ and confidence level $\delta$ and is
independent of $d$. 

The number of actual experiments that will have to
be performed on average, will, however, depend on the
system size, i.e., scale exponentially with the number of qubits, for
general unitaries. This is due to the $X_{\kappa_l}$ being
known only approximately and can be seen as follows:
The finite accuracy of the $X_{\kappa_l}$ gives rise to an
approximation of $F_{L}$,
$\tilde F_{L} = \frac{1}{L}\sum_{\kappa_l=1}^L \tilde X_{\kappa_l}$,
where the tilde indicates approximate values. Therefore, 
in addition to ensuring that  $F_{L}$ approximates $F_{av}^j$ with an
error of at most $\epsilon$, one also must guarantee that
$\tilde F_{L}$ approximates $F_{L}$ 
with the desired accuracy. This implies repeated measurements for a
given element $\kappa_l$ ($l=1,\ldots,L$) of the Monte Carlo sample.
Denoting the number of respective measurements by $N_l$, the total
number of experiments is given by $N_{exp}=\sum_{l=1}^L N_l$.
It can be shown~\cite{daSilvaPRL11,FlammiaPRL11} 
that choosing 
\begin{equation}
  \label{eq:Nl}
  N_l=\frac{1}{\epsilon L[\chi_U^{j}(\kappa_l)]^2}\log\left(\frac{4}{\delta}\right)  
\end{equation}
guarantees the approximations of $F_{L}$ by 
$\tilde F_{L}$ and of $F_{av}^j$ by $F_{L}$ to hold with the desired
confidence level. 

Since Monte Carlo estimation is carried out by randomly drawing $L$ times
an event from the $Td^2$-dimensional space of events, sampling
requires  $\mathcal C_{sampl}$ classical computational
resources. The sampling step is, for a general unitary, not efficient
since the dimension $d$ of the state space scales exponentially in the
number of qubits. Indeed, computing $\chi_U^{j}(i,k)$ 
requires to manipulate exponentially large matrices an exponential number
of times. Note that while 
the sampling procedure will select only some of the settings, the 
ability  to implement all of them is nevertheless required.
The total number of measurements $\langle N_{exp}\rangle$ that 
needs to be carried out on average is given by summing over
$N_l$ which in turn 
is inversely proportional to the weight of the setting in the 
relevance distribution, cf. Eq.~\eqref{eq:Nl}.
The scaling of resources required to estimate the average fidelity 
is therefore strictly connected  to the specific features of the
relevance distribution, or more specifically, of the characteristic
function $\chi_U^{j}(i,k)$ of the target unitary $U$ in the chosen
measurement basis $W_k$. If that  basis allows many
$\chi_U^{j}(i,k)$ to vanish and those that do not vanish to decrease
at most polynomially with the number of qubits, then the estimation
procedure is efficient.

\subsection{The special case of  Clifford qubit gates}
\label{subsec:Clifford}

Clifford gates acting on $n$ qubits are special in that they yield a
relevance distribution which has many zeros and all non-zero values are
identical. This in turn implies that the characterization of Clifford
operations is efficient, i.e., the average number of experiments is
independent of the number of qubits $n$ and the classical computational
effort scales only polynomially in $n$. 
In order to see why this is the case we briefly review  the definitions
of the Pauli group and the Clifford group as well as the action of
the Clifford group on Pauli measurements and  their eigenstates.  
Pauli observables, i.e., tensor products of single-qubit Pauli
operators, represent the natural measurements in the logical basis and
thus  constitute the standard measurement basis for $n$ qubits. 
This measurement basis can be considered 'minimal' in the sense that
it only assumes the ability of implementing single-qubit 
gates and readout  with no need for entangling operations~\footnote{
  Even though tensor products of single-qubit Pauli operators contain
  entangling operations, each Pauli operator can be measured in a separable
  eigenbasis.
}.

The set of Pauli measurements $\bar{\mathcal{P}}$ acting on $n$
qubits is defined as 
$\bar{\mathcal{P}}=\{\bar P_i=\bigotimes_{k=1}^n\sigma_{i_k}\}_{i=1}^{d^2}$
where each $\sigma_{i_k}$ represents a single-qubit Pauli operator
acting on the $k$th qubit, i.e., $i_k\in\{0,x,y,z\}$.
The operators in $\bar{\mathcal{P}}$ generate the Pauli group 
$\mathcal{P}=\{P_k=i^a\omega^b\bar P_j;\; 0<k\leq 4d^2\}$ with
$a,b=0,1$, $j=1,\ldots,d^2$, $\omega=\exp(i\pi)$ and matrix
multiplication being the group operation.
It is useful to introduce sets $\mathcal W_A$ of $d$ pairwise
commuting Pauli measurements.
For example, 
$\mathcal W_z$ comprises the $d$ different tensor products made up of
identities and $\sigma_z$'s.
 
The action of any transformation $U_C$ belonging to the Clifford
group is to map an element $P_i$ of $\mathcal{P}$ into 
another element $P_k$ of $\mathcal P$. In other words, the Clifford
group is the normalizer $\mathcal{N}(\mathcal P)$ of the Pauli group
in $U(d)$ since it leaves $\mathcal P$
invariant under conjugation. This implies for the 
orthonormal basis of Pauli measurements $\bar{\mathcal{P}}$ that each
element of $\bar{\mathcal{P}}$ is mapped into
another element from this set up to a phase factor, i.e., up 
to a permutation of eigenvalues~\cite{GottesmanPRA98}, 
\begin{equation}
  \label{eq:conjugation}
  U_C\bar P_kU_C^+ = \omega^a \bar P_i; \quad a=0,1. 
\end{equation}
Clifford operations can also be 
defined in terms of  their action on  
stabilizer states, i.e., in terms of  their action on the joint
eigenbasis of a set $\mathcal
W_A$~\cite{GottesmanPRA98,daSilvaPRL11}. 
One needs  to fix a particular eigenbasis  because each Pauli
measurement acting  on more than one qubit
is degenerate; and  it is thus not possible to characterize
the action of a Clifford
operation on a generic eigenbasis of a generic Pauli operator.
 Indeed, a Clifford operation maps joint eigenstates of the set
$\mathcal W_A$ into joint eigenstates of the set $\mathcal W_{A'}$, 
with either $A=A'$ or $A\neq
A'$~\cite{GottesmanPRA98,BermejoQuantInf14}. In general, one can
partition the set of Pauli measurements $\bar{\mathcal{P}}$ into $d+1$
commuting sets $\mathcal{W}_A$, i.e., $\bar{\mathcal{P}}$ exhibits the
so-called maximally partitioning property~\cite{LawrencePRA02}.  Each
partitioning defines a unique choice of $d+1$ joint eigenbases which
are mutually  unbiased with respect to each
other~\cite{BandyoAlgo02,LawrencePRA02}. 
The maximally partitioning property ensures that, if a state
$|\psi^A_i\rangle$ is a joint eigenvector of the operators in
$\mathcal{W}_A$, its expectation value vanishes for all Pauli
measurements outside of $\mathcal{W}_A$~\footnote{
The maximally partitioning property  also 
allows for an explicit construction of the $d+1$ MUB.
}. This can be seen as follows:
If the operator basis is maximally partitioning,
all operators outside of $\mathcal{W}_A$ can be expressed
in terms of an eigenbasis which is mutually unbiased with
respect to $\{\ket{\psi^A_i}\}$. 
We recall that two complete and orthonormal bases $A$, $A^\prime$ on a
$d$-dimensional Hilbert space are mutually unbiased if and only if
\begin{equation}
  \label{eq:MUB}
  |\langle\psi^A_i|\psi^{A^\prime}_j\rangle|=1/\sqrt{d}  
\end{equation}
for all $|\psi^A_i\rangle\in A$, $|\psi^{A^\prime}_i\rangle\in
A^\prime$~\cite{WoottersAnnPhys89}. 
For a generic Pauli measurement belonging to the commuting set
$\mathcal{W}_{A^\prime}$, $\bar P_k=
\sum_l\lambda^k_l|\psi^{A^\prime}_l\rangle\langle\psi^{A^\prime}_l|$,  
the expectation value is given by 
\[
 \Tr\left[\bar P_k\ket{\psi^A_i}\bra{\psi^A_i}\right] = 
 \sum_{j,l=1}^d\lambda^k_l
 |\langle\psi^A_i|\psi^{A^\prime}_l\rangle|^2.
\]
If $\mathcal{W}_A\neq \mathcal{W}_{B}$, then
$|\langle\psi^A_i|\psi^{A^\prime}_j\rangle|^2=1/d$ 
and 
\[
\Tr\left[\bar P_k\ket{\psi^A_i}\bra{\psi^A_i}\right] =
\frac{1}{d} \sum_{l=1}^d\lambda^k_l=0
\]
since Pauli measurements are traceless. Therefore 
\begin{equation}\label{eq:mubmeas}
 \Tr\left[\bar P_k\ket{\psi^A_i}\bra{\psi^A_i}\right] = 
 \begin{cases} 
   \omega^a & \mbox{if}\;\; \bar P_{k}\in\mathcal{W}_A\\
   0&\mbox{otherwise}
 \end{cases} \,.
\end{equation}
Equation~\eqref{eq:mubmeas} is a consequence of the fact 
that measurements associated to MUB span orthogonal
subspaces~\cite{WoottersAnnPhys89}. 

In the context of Monte Carlo estimation of the average gate fidelity
for a Clifford gate,  
Eq.~\eqref{eq:mubmeas} gives rise to a uniform relevance distribution. 
In order to elucidate this, we distinguish whether the set  of
inputs $I$ is made up of states (belonging to MUB)~\cite{ReichPRL13}
or operators~\cite{FlammiaPRL11,daSilvaPRL11}. In the
former case, applying Eq.~\eqref{eq:mubmeas} to each state 
$|\psi^A_i\rangle\langle\psi_i^A|\in I$ yields for the characteristic
function, cf. Eq.~\eqref{eq:chi_U},
\begin{eqnarray}\label{eq:cliffchist}
  \chi^{j}_{U_{C}}(i,k) &=&
  \Tr\left[\bar P_k U_{C}\ket{\psi^A_i}\bra{\psi^A_i}U_C^+\right]= 
  \Tr\left[\bar P_k\ket{\psi^{A^\prime}_{m}}\bra{\psi^{A^\prime}_{m}}\right] \nonumber
    \\ &=&
 \begin{cases} 
   \omega^a & \mbox{if}\;\; \bar P_{k}\in\mathcal{W}_{A^\prime}\\
   0&\mbox{otherwise}
 \end{cases} \,,
\end{eqnarray}
where $\ket{\psi^{A^\prime}_m}$ is the $m$th element of the joint 
eigenbasis of the  commuting set $\mathcal{W}_{A^\prime}$.
Inserting this into Eq.~\eqref{eq:rel} leads to 
\begin{equation}
  \label{eq:cliffstates}
  P^{j}_{U_C}(i,k)=\frac{1}{\mathcal{N}}\,,
\end{equation}
i.e., the relevance distribution is uniform. It contains
$\mathcal{N}=Td$ non-zero elements 
since for each of the $T$ input states there are only $d$ non-vanishing
measurements.  Sampling then simply amounts to randomly drawing an
index $i\in[1,T]$ to select the input state and, after 
calculating the output state from the action of the  Clifford
operation on the input state,  
to randomly draw an index $k\in[1,d]$ to select
the output measurement from the commuting set corresponding
to the output state. Uniformity of the relevance distribution implies
that the sampling is independent of  system size such that 
$\mathcal{C}_{sampl}\propto\mathcal{O}(1)$. Following the
Gottesman-Knill theorem for Clifford circuits~\cite{GottesmanKnill}, 
the  overall classical computational resources to
calculate the output state scale polynomially in $n$. 

If the  set $I$ of inputs is made up of operators,
one can directly use the definition of the 
Clifford group as the normalizer of the 
Pauli group, Eq. \eqref{eq:conjugation},
to obtain
\begin{eqnarray}\label{eq:cliffchiop}
  \chi^{j}_{\mathcal{U}_C}(i,k) &=& 
 \frac{1}{d} \Tr\left[\bar P_k\,\mathcal{U}_C\left(\bar P_i\right)\right]=
  \frac{1}{d}\Tr\left[\bar P_k U_C\bar P_i U_C^+\right]\nonumber \\
  &=&\frac{\omega^a}{d} \Tr\left[\bar P_kP_{k'}\right] =
  \pm  \delta_{kk'}\,.
\end{eqnarray}
Together with Eq.~\eqref{eq:rel}, this leads to 
\begin{equation}
  P^{j}_{U_C}(i,k)=\frac{1}{\mathcal{N}}
\end{equation}
with $\mathcal{N}=d^2$.
For each input operator there is only one output which leads to
a non-zero outcome. Sampling amounts to randomly drawing an index
$k\in[1,d^2]$ and finding $i$ such that $\pm\bar P_i=U_C\bar P_kU_C^+$.
The latter can be done 
efficiently on a classical computer due to the Gottesman-Knill
theorem~\cite{GottesmanKnill}. 
Once the pair of input operator/output measurement has been
selected, a second sampling step is required to randomly draw 
an eigenstate of the input operator $\bar P_k$.
This step is computationally efficient since the spectrum
of each operator corresponds to a uniform distribution.
As a result, the sampling complexity  $\mathcal{C}_{sampl}$ 
is independent of system size and the classical
computational resources  scale polynomially
in $n$ also for input operators~\cite{daSilvaPRL11}.

The number of non-zero elements of the relevance
distribution for a Clifford gate is either $Td= \mathcal{N}$, for the
protocols based on input states, or $d^2=\mathcal N$ for the
entanglement fidelity protocol, as opposed to $T d^2$ for a generic 
unitary, independent of the protocol. This 
implies  efficient scaling of the average number of
experiments $\langle N_{exp}\rangle$ that have to be carried out for
Clifford gates.
In general, $\langle N_{exp}\rangle$ can be estimated
by averaging over the  number $N_l$ of repetitions for each setting
with the weights in the averaging given by the probability distribution
$P^j(i_l,k_l)$~\cite{daSilvaPRL11,FlammiaPRL11,ReichPRL13}. 
For a generic unitary, this yields 
\begin{eqnarray}\label{eq:expscal}
  \langle N_{exp}\rangle &=&
  \sum_{i_l=1}^{T}\sum_{k_l=1}^{d^2}P^{j}(i_l,k_l)N_l\nonumber\\
  &=&\frac{1}{\mathcal{N}}\sum_{i_l=1}^{T}\sum_{k_l=1}^{d^2}
  \left[\chi^{j}(i_l,k_l)\right]^2
  \frac{4}{\left[\chi^{j}(i_l,k_l)\right]^2L\epsilon^2}
  \log\left(\frac{2}{\delta}\right) \nonumber\\
  &\propto&\frac{1}{\mathcal{N}}Td^2
  =
  \begin{cases}
    \mathcal{O}(d^2) & \mbox{for}\;\;\mbox{operator inputs} \\
   \mathcal{O}(d) &  \mbox{for} \;\;\mbox{state inputs}    
  \end{cases}\,.
\end{eqnarray}
The scaling is obtained from observing that 
$\kappa_l=(i_l,k_l)$ can take $Td^2$ values whereas 
$\mathcal N=d^2$ for operator inputs and $\mathcal N=Td$ for state
inputs and $T=d^2$ for operator inputs. For Clifford gates, 
due to Eq.~\eqref{eq:cliffchist}, respectively,
Eq.~\eqref{eq:cliffchiop}, this reduces to  
\begin{equation}\label{eq:cliffexpscal}
  \langle N_{exp}\rangle 
  \propto\frac{1}{\mathcal{N}}\mathcal{N}=\mathcal{O}(1)\,.
\end{equation}
The fact that the number of experiments that need to be carried out is
independent of system size implies that estimating the average gate
fidelity is maximally efficient.

\section{Operator bases for qudits}
\label{sec:opbases}

The discussion in the previous section suggests 
that the existence of a class of unitaries for which $F_{av}$
can be estimated with maximal efficiency is due to two fundamental
ingredients: (i) existence of a non-trivial class of unitaries 
$(\mathbb{U}_C=\{U_j\neq\openone\})$ which
map the operator basis into itself, up to a phase-factor; 
(ii) uniformity of the associated relevance distribution. 
Condition (i) implies that the relevance
distribution associated to this class of unitaries contains  a 
reduced number $\mathcal{N}$ of non-zero elements which leads to 
$\langle N_{exp}\rangle \propto\mathcal{O}(1)$. Condition
(ii) ensures that also the sampling step is efficient since
the coefficients of the relevance distribution
are known \textit{a priori} with no need of explicit calculation. 
Both these features are intimately related to the properties of the  
Pauli measurement basis. 

Specifically, they are connected to 
the fact that the set of the standard Pauli measurements can be
partitioned into $d+1$ commuting sets. This can be seen as 
follows: As shown in the previous section, condition (ii) 
follows from Eq.~\eqref{eq:mubmeas} which in turn results from the
standard Pauli measurements being associated to MUB that span
orthogonal subspaces, i.e., from the Pauli measurements allowing for a 
maximal partitioning. It seems highly likely that 
the maximally partitioning property is also  
a necessary condition for (i), i.e., the existence of target unitaries
which map the measurement basis into itself, up to a phase
factor. 
The close connection between the maximally partitioning property
and the existence of $\mathbb U_C$ can be inferred from the fact that
Clifford operations can be defined as those unitaries that map
stabilizer states into stabilizer states. That is,
ensuring the existence of  $\mathbb U_C$ corresponds to  ensuring the  
existence of generalized stabilizer states. These are the common
eigenstates of $d$ pairwise commuting measurement operators 
that have a non-vanishing expectation value only on this set of
operators. In other words, the generalized stabilizer states are 
mutually unbiased joint eigenstates.
The maximally partitioning property by itself is, however, not sufficient 
to ensure efficient characterization. Additionally, 
the spectra of the measurement operators must obey certain constraints.
The dependence of the relevance distribution on the spectrum of the
basis operators is apparent from Eqs.~\eqref{eq:cliffchist}
and~\eqref{eq:cliffchiop}.

In order to determine whether there exist qudit operations that can be
efficiently characterized, we thus need to find 
a suitable generalization of the Pauli measurements. Since Clifford
gates are defined in terms of the measurement basis,
cf. Eq.~\eqref{eq:conjugation}, this implies also identification of
the class of unitaries $\mathbb{U}_C$ that corresponds to the specific
choice of measurement basis. Unfortunately,
it is not possible to generalize all properties of the standard Pauli
measurements for qubits to higher dimensions. 
Most notably, for $d>2$ and $d\neq 2^m$ with $m$ a positive integer, 
unitarity and Hermicity cannot  be enforced at the same time  on an
orthonormal and complete operator basis. Hence, when replacing qubits
by qudits, it is crucial to understand what are the properties of the
standard Pauli measurements that the generalized operator basis must 
retain for efficient estimation of the average fidelity. 
Moreover, it is important to determine how  different features of the
operator basis affect the scaling of resources of the Monte Carlo
procedure. For the latter, we distinguish between the average number
$\langle N_{exp}\rangle$  of experiments and the classical
computational resources $\mathcal C_{sampl}$ needed for the
sampling. $\langle N_{exp}\rangle$ 
becomes independent of system size if the relevance distribution has
the minimal number of non-zero elements,
cf. Eq.~\eqref{eq:cliffexpscal}. Efficient sampling in the standard
MC approach requires in 
addition that the relevance distribution is uniform.

To identify the generalized measurement operators and the associated
unitaries that leave it invariant under conjugation, we start from
what we argue to be  the
fundamental requirement for efficient characterization -- existence of
$d+1$ MUB. Since they are the joint eigenbases of the measurement
operators in the commuting sets of the maximally partitioning basis, the
unitaries that map the operator basis onto itself should
also map  the set of $d+1$ MUB into itself.
We utilize this property to 
determine candidates for the class of unitaries that can be 
characterized efficiently in Sec.~\ref{subsec:mub}. In particular, we
show that any change of basis between two bases in the set of MUB leaves
this set invariant.
In Sec.~\ref{subsec:maxpart} we discuss the construction of an
operator basis out of the $d+1$ MUB and the difficulty of guaranteeing
the maximal partitioning property for the operator basis.
We therefore distinguish between the single qudit and multiple qudit
cases and impose the conditions for $\langle
N_{exp}\rangle\propto\mathcal O(1)$ at the single qudit level in
Sec.~\ref{subsec:maxpart_single}. This ensures the average number of
experiments to be independent of system size for those $U_j\in
\mathbb{U}_C$ that are given by tensor products of single qudit
unitaries. The conditions allow for both unitary and Hermitian
operator bases. 
In order for $\mathbb{U}_C$ to also comprise
entangling operations, we need to impose the conditions for $\langle
N_{exp}\rangle\propto\mathcal O(1)$ at level of multiple qudits in
Sec.~\ref{subsec:maxpart_multiple}. These conditions also allow for
both unitary and Hermitian operator bases. However, it is not clear if
a Hermitian basis satisfying these constraints will correspond to
local measurements, whereby we mean those measurements that can be
expressed as tensor products of   
single-qudit operators. Most likely this is not the case.

We continue with the conditions for efficient sampling in
Sec.~\ref{subsec:unif} and show that in order to ensure a uniform
relevance distribution, the spectrum of the measurement operators must
be made up of roots of unity and zero. This together with the requirement
for the operator basis to be orthonormal and traceless rules out
Hermitian operators. In contrast, a unitary operator basis not only
allows for efficient sampling but also maximizes the set
$\mathbb{U}_C$ and can be constructed in terms of local
measurements. Clearly, the notion of unitary, non-Hermitian
measurements is non-standard. We therefore discuss the experimental
implementation of such measurements in Sec.~\ref{subsec:genpauli}.
 
\subsection{Unitary transformations between two MUB}
\label{subsec:mub}

We denote the set of $d+1$ MUB by $\mathcal{M}$.
Since, on a $d$-dimensional Hilbert space, $d+1$ MUB 
are guaranteed to exist only
if $d$ is equal to a prime number or a power of a prime
number~\cite{WoottersAnnPhys89}, we restrict our 
investigation to $p$-level systems with  $p$ prime (qupits).
We examine the properties of unitary transformations that 
map two bases in $\mathcal M$ into each other. In particular,
any such transformation is  a mapping
from $\mathcal{M}$ into itself. Or more formally: 

\textbf{Proposition 1:} 
Consider 
a basis $A_j\in\mathcal M$, $j\in[1,d+1]$, with elements
$\ket{\psi_k^j}$, $k\in[1,d]$. Any unitary transformation
between the elements of $A_j$ and $A_{j'}\in\mathcal{M}$,
\begin{equation}\label{eq:ugen}
  U_{jj^\prime}=\sum_{k=1}^d\ket{\psi^{j^\prime}_{k}}\bra{\psi^j_k}\,,
\end{equation}
will also be a unitary transformation between the elements of
 $A_i\in\mathcal{M}$ and $A_{i'}\in\mathcal{M}$
with $i^\prime=i+(j-j^\prime)]$ for each $i\in[1,d+1]$ and the sum
modulo $d+1$
 i.e., 
\begin{equation}\label{eq:ugenbis}
  U_{jj^\prime}=U_{\delta}\,,
\end{equation}
with $\delta=j'-j$.

We prove Proposition~1 in Appendix~\ref{subsec:proof_prop1} and provide
here an intuitive interpretation. Visualizing the indices $j$ of the
bases in $\mathcal{M}$ geometrically  as points on a line, mutual
unbiasedness implies that all points are equally
spaced and therefore must lie on a circle.
The freedom in the phase factor of the overlap between two elements of
two MUB, cf. Eq.~\eqref{eq:MUB}, accounts for the 
number of steps separating the  points   on the circle.
Given such a regular structure,
any transformation which maps basis $A_j\in\mathcal{M}$ 
into basis $A_{j^\prime}\in\mathcal{M}$ can be interpreted as
a shift of $j^\prime-j$ steps on the circle,
regardless of the starting point. Hence, any shift of $\delta$ 
steps on the circle corresponds to a mapping, 
modulo $d+1$, between any two bases in $\mathcal M$
whose corresponding indices $i,i^\prime$ are
$\delta$ steps apart.

We show in Appendix~\ref{subsec:Ugen_group} 
that the unitaries defined by Eq.~\eqref{eq:ugenbis} 
can be decomposed into a transformation
$U^0_\delta$, which maps the $k$th element
of basis $A_i$ into the $k$th element of $A_{i+\delta}$,
and a permutation of the elements of the two bases.
We then prove in Appendix~\ref{subsec:Ugen_group} 
that the unitaries  defined by Eq.~\eqref{eq:ugenbis} 
form a group under matrix multiplication, 
$\mathbb{U}^\Pi_\delta=\{U_\delta\}$.
It can be interpreted as the composition
of the group of permutations with the group
of transformations $\mathbb{U}^0_\delta=\{U^0_\delta\}$.
The unitaries in $\mathbb{U}^\Pi_\delta$ are the
candidates for $\mathbb{U}_C$, hence for
efficient characterization, once an
operator basis is constructed from the MUB.

\subsection{Maximally partitioning operator basis} 
\label{subsec:maxpart}

Given a set $\mathcal M$ of $d+1$ MUB, an operator basis can be
constructed in terms of projectors onto the states of the MUB. 
This operator basis  is, by construction,
 maximally partitioning. We
recall the formal definition of a maximally
partitioning operator basis~\cite{BandyoAlgo02,LawrencePRA02}:

\textbf{Definition:} An orthonormal and complete
operator basis $\mathbb{B}$ on a $d$-dimensional
Hilbert space is maximally partitioning if there exists
a $d+1$-dimensional set $\mathcal{M}=\{A_j\}_{j=1}^{d+1}$ of 
mutually unbiased bases $A_j=\{\ket{\psi^j_k}\}_{k=1}^d$
such that every operator in $\mathbb{B}$
can be expressed as
\begin{equation}\label{eq:specdecomp}
  B^{j}_i=\sum_{k=1}^d\lambda^j_{i,k}\ket{\psi^j_k}\bra{\psi^j_k}\,.
 \end{equation}
In Eq.~\eqref{eq:specdecomp}, 
$\lambda^j$ is a  $d\times d$ matrix
whose rows are orthogonal. Each entry  $\lambda^j_{ik}$ corresponds to
the $k$th eigenvalue of the $i$th operator in $\mathbb{B}$
 sharing the eigenbasis  $\{\ket{\psi^j}\}$, i.e., belonging to 
 the commuting set $\mathcal{W}_j$.
In particular, since the first row of each $\lambda^j$ corresponds
to the spectrum of the identity,  
$\sum_{k=1}^d\lambda^j_{i,k}=0$ for each $j\in[1,d+1]$ and $i\in[2,d]$.

The identity needs to be included in the operator basis since it is
left invariant by any unitary transformation and is diagonal in each
of the bases in $\mathcal{M}$.
Orthogonality of the rows of $\lambda^j$ guarantees
orthonormality of the operators within the same 
commuting set. The condition $\sum_{k=1}^d\lambda^j_{i,k}=0$ ensures 
that all operators are orthogonal to the identity as well as 
that operators in different commuting sets are orthogonal.

In practical device characterization, the measurement operators should
be tensor products of single-qupit operators. Then the measurements
are local in the sense that each operator can be measured in a
separable eigenbasis. 
The construction of an operator basis from the MUB which obeys the
tensor product structure is far from
straightforward. The proof of Ref.~\cite{WoottersAnnPhys89} 
ensures existence of the set of MUB but does not provide a
prescription on how to actually construct the corresponding
observables. For unitary operators, 
such a prescription is found in Ref.~\cite{BandyoAlgo02} starting from
a maximally partitioning basis for a single qupit:
It can be shown that the maximally partitioning  property is preserved
under the tensor product by making explicit use of unitarity of the
single-qupit operator basis. 
The  maximally partitioning basis for multiple qupits is then 
obtained by tensor products of the single-qupit unitary basis
operators~\cite{BandyoAlgo02,LawrencePRA11}. 
A weaker version of this result holds also for other maximally
partitioning bases, for example Hermitian ones:
Given the spectral decomposition~\eqref{eq:specdecomp}, 
the $\lambda^j$ matrices for multiple qupits can be constructed as
tensor products of the $\lambda^j$ matrices for $n=1$ 
since orthonormality and completeness of the operator
basis are preserved under tensor product. However, this does not
ensure that the maximally partitioning operator basis itself
can be constructed as tensor products of the single-qupit
operators. In general, that is, without making any assumption on the
spectra of basis operators, one obtains only $p+1$ out of the
$p^n+1$ bases in $\mathcal{M}$ by tensor products.
This is not enough to ensure a maximal partitioning for the resulting
operator basis. While it seems reasonable to expect that the
maximally partitioning property is preserved only for unitary
operators, it remains an open question whether this holds also for 
an Hermitian operator bases and if so, under which spectral
conditions. 

We therefore distinguish between imposing the maximally partitioning
property at the single  at the  multi-qupit level. If
only the single-qupit operator basis needs to give rise to a maximal
partitioning, the multi-qupit operator basis which is constructed by 
tensor products from the single-qupit basis is not guaranteed to
inherit this property.
This implies that only unitaries that are themselves tensor products,
i.e., non-entangling operations,  yield $\langle
N_{exp}\rangle\propto\mathcal O(1)$.  

\subsection{Ensuring $\langle N_{exp}\rangle\propto \mathcal O(1)$
  at the single qupit level}
\label{subsec:maxpart_single}

The average number of experiments required to characterize
a unitary transformation, $\langle N_{exp}\rangle$, becomes
independent of system size if the 
relevance distribution has a reduced number, $\mathcal N$,  of
non-zero elements.  
We now determine the corresponding conditions on the operator
basis $\mathbb{B}$. To
differentiate between single and multiple qupits, we indicate the
dependence of the operator basis on the number $n$ of qupits by
$\mathbb{B}(n)=\{B_i(n)\}_{i=1}^{d^2}$ where  $d=p^n$, $n\geq1$. 
 Analogously for the group of unitaries
 that leaves $\mathbb{B}(n)$ invariant. 
The conditions at the single qupit level are given by 

\textbf{Theorem 1:} 
For any $n$, a non-trivial class of unitaries, $\mathbb{U}_C(n)$,
for which $\langle N_{exp}\rangle\propto \mathcal O(1)$ exists if
\begin{enumerate}
\item the operator basis for a single qupit, $\mathbb{B}(1)$, 
  is  maximally partitioning, 
\item all single-qupit $\lambda^j$'s in the decomposition
  \eqref{eq:specdecomp} are equal. 
\end{enumerate}

We prove this theorem in Appendix~\ref{subsec:theorem1}. 
The idea underlying the proof 
is the following: Conditions 1 and 2 ensure 
that the single-qupit operator basis $\mathbb{B}(1)$
is left invariant by the group of transformations 
$\mathbb{U}^0_\delta(1)$. 
Consider the multiple-qupit operator basis $\mathbb{B}(n)$ that is 
obtained from tensor products of the elements  of $\mathbb{B}(1)$.
By construction it  is
left invariant by the set of transformations $\tilde{\mathbb{U}}^0_\delta(n)$, 
obtained from tensor products of the elements of  $\mathbb{U}^0_\delta(1)$. 
The transformations  in $\tilde{\mathbb{U}}^0_\delta(n)$ obey a relation analogous
to Eq.~\eqref{eq:cliffchiop} and  thus
yield an average number of experiments that is independent of system
size for the protocol based on the entanglement fidelity.

By construction, the operators in $\mathbb{B}(n)$ admit the existence
of a set $\mathcal{M}^{sep}(n)$ of $p+1$ separable mutually unbiased joint
eigenbases. These are obtained from tensor products of the elements of
the single-qupit set of MUB, $\mathcal{M}(1)$. The set $\mathcal{M}^{sep}(n)$
is mapped into itself by the transformations
in $\tilde{\mathbb{U}}^0_\delta(n)$, and the states belonging to it
obey a relation analogous  to Eq. \eqref{eq:mubmeas}.
Hence, 
if the characterization protocol does not require more than
$p+1$ MUBs,  the relevance distribution of the transformations in 
$\tilde{\mathbb{U}}^0_\delta(n)$ has $\mathcal{N}$ non-zero elements
and $\langle N_{exp}\rangle\propto\mathcal{O}(1)$. 
This is the case for the protocol based on the two classical fidelities 
but not for the two-design protocol. 
Since the latter requires $d+1$ MUB,  it can not be used in
combination with an operator basis that only ensures existence of
$p+1$ MUB.

Conditions 1 and 2 thus ensure the existence 
of a group of unitaries, $\tilde{\mathbb{U}}^0_\delta(n)$, that 
lead to $\langle N_{exp}\rangle\propto\mathcal{O}(1)$
for the protocol sampling over the eigenstates of input
operators~\cite{daSilvaPRL11,FlammiaPRL11}  
and for the protocol based on the two classical fidelities~\cite{ReichPRL13}.
The group of transformations $\tilde{\mathbb{U}}^0_\delta(n)$ 
represents only  a subgroup of the group
$\mathbb{U}^0_\delta(n)$ since the latter must also contain
entangling operations, i.e., operations which cannot be obtained as
tensor products of single-qupit unitaries. This follows from the proof
of Ref.~\cite{LawrencePRA11} showing that, for $n>1$, bases with a
different entanglement structure coexist
within the same set of MUB $\mathcal{M}(n)$. Therefore
the group  $\mathbb{U}^0_\delta(n)$ includes transformations
mapping two bases with a different entanglement content into  
each other, i.e., entangling operations. 

Theorem~1 is compatible with both real 
and  complex spectra of the measurement operators, i.e., with 
both  unitary and Hermitian operator bases. However, 
for qutrits ($p=3$),  the Gell-Mann basis, i.e., the basis of 
the standard generators of $SU(3)$, does not fulfill the
conditions of Theorem~1 since the 
eigenvectors of the Gell-Mann operators are not  mutually unbiased. 
This also implies that such a basis cannot be 
used with the two protocols based on input states~\cite{ReichPRL13}
for any unitary. 

The conditions in Theorem~1 define a minimal underlying regular
structure of the operator basis. 
We assume that in absence of such a regular structure
it is not possible to find any transformation, besides the identity,
that maps the full operator basis into itself.
Conditions 1 and 2 then endow a generic operator basis with the most
general regular structure it can have that allows for 
a relevance distribution with a reduced number of elements
at least in some protocol, at least for some unitaries. 

\subsection{Ensuring $\langle N_{exp}\rangle\propto \mathcal O(1)$
  at the level of $n$ qupits}
\label{subsec:maxpart_multiple}

In order to achieve a number of experiments that is independent of
system size for any
number of qudits and independent of the protocol, 
the operator basis needs to allow for a  maximal partitioning
for every $n$. This is expressed by

\textbf{Theorem 2:}
A non-trivial class of unitaries, $\mathbb{U}_C$, for which
the scaling of $\langle N_{exp}\rangle$ is
$\mathcal{O}(1)$, independent of the characterization 
protocol, exists if
\begin{enumerate}
\item the operator basis $\mathbb{B}(n)$ is  maximally partitioning, 
\item all $\lambda^j$'s in the decomposition~\eqref{eq:specdecomp} are equal.
\end{enumerate}
This class of unitaries includes entangling operations.

This theorem can be proven in exactly the same way as  
for single qupits in Appendix~\ref{subsec:theorem1}, i.e., by
substituting the single-qupit operators in Eq.~\eqref{eq:theo1} by 
multi-qupit operators.
Assuming the operator basis to be maximally partitioning for every $n$
ensures that one can construct the full set $\mathcal{M}(n)$ of MUB
out of the joint eigenbases of the operators in $\mathbb{B}(n)$. 
This implies  that $\mathbb{U}^0_\delta(n)\subseteq\mathbb{U}_C(n)$
for all protocols. 
The set $\mathbb{U}^0_\delta(n)$ includes entangling operations 
since the MUB in $\mathcal{M}(n)$
have different entanglement content~\cite{LawrencePRA11}.

Theorem~2 is compatible with both  real and complex spectra of
the measurement operators. 
However, it might not be possible to obtain
a Hermitian basis from tensor products of the single-qupit bases which
allows for a maximal partitioning. In that case, the Hermitian
operators would not correspond to local measurements. 
For an operator basis that gives rise to a maximal partition and is
constructed in terms of tensor products of single-qupit operators, 
condition~2 of Theorem~2 translates
into the requirement that 
all single-qupit operators have the same spectrum.

So far we have identified a set of conditions that guarantee
the average number of experiments in Monte Carlo estimation of
$F_{avg}$ to be independent of system size, 
$\langle N_{exp}\rangle \propto\mathcal O(1)$, for certain unitaries. This
corresponds to a first step towards efficient Monte Carlo
characterization. The additional condition of a uniform relevance
distribution, that ensures the classical computational resources to scale
at most polynomially in $n$, requires additional constraints on the 
spectra of the measurement operators. 

\subsection{Ensuring efficient sampling: The uniform relevance
  distribution} 
\label{subsec:unif}

Efficient sampling requires a uniform relevance distribution which
together with tracelessness and orthonormality
of the operator basis, implies the measurement 
operators to have the same spectrum, up to a phase factor, with 
the modulus square of each eigenvalue being equal to 1.
For Hermitian operators, uniformity of the relevance distribution
combined  with the constraint
of tracelessness, implies that the spectrum of each basis
operator must be the same and 
made up of an equal number of $+1$ and $-1$ and at least 
one zero. However, for $p>2$, 
such a spectrum is incompatible with orthonormality of the operator
basis. It is easy to check that already for a single qutrit ($p=3$),
this choice of eigenvalues does not allow to construct
a $(p\times p)$-matrix $\lambda$ with orthogonal rows.  This holds also
for prime numbers $p>3$. 
As a consequence, enforcing the operator
basis to be Hermitian rules out the possibility of obtaining
a uniform relevance distribution and thus efficient sampling 
for any target unitary (except identity).

In contrast, for unitary measurement bases, tracelessness and
unitarity  imply that the spectrum of each single-qupit operator is
$p$-nary, i.e., made up of the $p$ distinct $p$th roots of unity.
Consequently  the spectra of all multi-qupit
measurement operators are identical since each
$p$th root of the identity simply appears with multiplicity $p^{n-1}$.
Such a spectrum is also compatible with orthonormality.
Indeed, using $p$ distinct $p$th roots of unity, 
one can construct, for each of  the $p+1$ bases in $\mathcal{M}(1)$,
a set of  exactly $p-1$ pairwise orthogonal traceless
operators, i.e., 
a maximally partitioning single-qupit basis~\cite{Daniel14}. 
Since the maximal partitioning is preserved under tensor
product~\cite{BandyoAlgo02}, a $p$-nary spectrum is also compatible
with  a multiple-qupit 
operator basis that gives rise to a maximal partitioning. 
As a consequence,
a maximally partitioning unitary basis is compatible with a uniform
relevance distribution.  
It requires, however, a generalization of 
the relevance distribution given in Eq.~\eqref{eq:rel} 
to include complex expectation values, 
\begin{equation}\label{eq:relcompl}
P^{j}(i,k)=\frac{1}{\mathcal{N}} \left|\chi_U^{j}(i,k)\right|^2; \quad
\chi_U^{j}(i,k)\in\mathbb{C}\,.
\end{equation}
More formally, the conditions on the spectrum can be expressed as follows.

\textbf{Theorem 3:} 
A non-trivial set of unitaries $\mathbb{U}_C$ that can be
characterized efficiently  both in terms of the average
number of experiments and the classical computational
resources for any number of qupits  exists if
the single-qupit operator basis is maximally partitioning and unitary.

This theorem can be proven straightforwardly from the previous discussion: 
Since the maximally partitioning unitary basis satisfies conditions~1
and 2 of Theorem~2, then the set of transformations $\mathbb{U}_C$
which allows for efficient characterization contains at least
$\mathbb{U}^0_\delta(n)$ and therefore is non-trivial.
Moreover, due to the unitary spectrum of the basis operators,
the operations in  $\mathbb{U}^0_\delta(n)$
satisfy Eq.~\eqref{eq:mubmeas}
with $\omega=\exp(2i\pi/p)$ and  $a\in[0,p-1]$. This leads to 
a  generalized uniform relevance distribution, Eq.~\eqref{eq:relcompl},
in all protocols. We show below in Sec.~\ref{sec:qudit_reldist} that
such a generalized uniform relevance distribution 
yields $\mathcal{C}_{sampl}\propto\mathcal{O}(1)$.

For a maximally partitioning unitary operator basis,
the set of unitaries which leave the basis invariant up to a phase factor 
is larger than $\mathbb{U}^0_\delta(n)$.
This can be inferred from the fact that the operator basis is left
invariant, up to a phase factor, also by  arbitrary cyclic permutations 
and those permutations which map basis operators belonging to the same
commuting set into each other~\cite{Daniel14}.
Most likely, the set of unitaries given by $\mathbb{U}^0_\delta(n)$
extended by those permutations is also maximal.
However, whether this is indeed the case and whether the set 
coincides with the full group $\mathbb{U}^\Pi_\delta(n)$
of transformations which leaves the set $\mathcal{M}(n)$ of MUB
invariant remains an open question.

\subsection{A unitary operator basis vs actual measurements: The
  generalized Pauli basis} 
\label{subsec:genpauli}

A maximally partioning unitary operator basis is the so-called
generalized Pauli  basis~\cite{BandyoAlgo02,LawrencePRA02,GottesmanChaos99}.
This basis generates a group under matrix multiplication, the
generalized Pauli group. The  
group of transformations,  $\mathbb{U}_C(n)$,
 leaving the operator basis invariant
up to a phase factor, 
can be identified with the normalizer of the generalized Pauli group, i.e., 
with the generalized Clifford group~\cite{GottesmanChaos99}.
To construct the generalized Pauli basis, one generalizes
the standard Pauli $\sigma_z$
and $\sigma_x$ operators~\cite{BandyoAlgo02,LawrencePRA02,GottesmanChaos99},
\begin{eqnarray}
  Z (1)&=& \omega^n\ket{n}\bra{n}\,,\nonumber\\
  X (1)&=&\ket{n+1}\bra{n}\,,\label{eq:zx}
\end{eqnarray}
where addition is modulo $p$, $n\in[0,p\!-\!1]$, and $\omega=\exp{(2i\pi/p)}$.
The generalized Pauli operator basis for a single qupit
is obtained as~\cite{GottesmanPRA01} 
\begin{equation}
  X^a(1)Z^b(1)\quad a, b=0,\cdots,p-1\label{eq:pauliqud}\,.
\end{equation} 
For example, by setting $Y(1)=X(1)Z(1)$ and $V(1)=X(1)Z(1)^2$ the full operator
basis for a single qutrit reads $\bar{\mathcal{P}}(1)=\{I(1),X(1),Y(1),Z(1),V(1),X^2(1),Y^2(1),Z^2(1),V^2(1)\}$.
Each operator from the set commutes only with itself, its square
(corresponding to both its Hermitian conjugate and its inverse) and
the identity, i.e.,  with the operators obtained from a special 
set of  permutations identified in Ref.~\cite{Daniel14}. 
This defines for qutrits a unique partitioning into $d+1=4$ 
sets of commuting operators. The generalized Pauli basis,
Eq.~\eqref{eq:pauliqud}, gives rise to the definition of 
the generalized single-qupit Pauli group
as~\cite{LawrencePRA04,GottesmanChaos99} 
\begin{equation}\label{eq:pauligroup}
  \mathcal{P}(1)=\{\omega^iX^a(1)Z^b(1)\quad a, b,i\in[0,p-1]\}\,.
\end{equation}
In analogy to the qubit case, the Pauli measurements on $n$
qupits are given by tensor products of the single-qupit operators,
Eq. \eqref{eq:pauliqud}, 
which are also the generators of the $n$-qupit Pauli group.

To summarize, by enforcing unitarity on the $\lambda$-matrix
in Eq.~\eqref{eq:specdecomp}, we can obtain an operator basis which
generalizes all the fundamental properties
of the standard Pauli operators besides Hermicity. That is, an
orthonormal basis of unitary operators with a maximal partitioning
into $d+1$ commuting subsets which is preserved under tensor product. 
The $p$-nary spectrum of the basis is
preserved as the number of particles increases, 
and the operator basis generates a group under matrix
multiplication. Since we can define a generalized Clifford group and
obtain a uniform relevance distribution,  
the fundamental requirements for achieving  efficient
characterization for certain unitaries are met. 
There are two caveats,
however: (i) The Monte Carlo procedure needs to be generalized for
measurement operators with complex eigenvalues. This is done in
Appendix~\ref{sec:MC_complex}. (ii) Observables have to be
Hermitian, so we need to clarify how a unitary, non-Hermitian
measurement basis can be connected to measurable observables. There
are two options -- one can construct Hermitian counterparts of unitary
basis operators or utilize the concept of a quantum circuit to
simulate a Hermitian measurement.

A Hermitian counterpart can be constructed from the unitary
orthonormal set of generalized Pauli operators   
$\bar{\mathcal{P}}(1)=\{U_k(1)\}_{k=1}^{p^2}$ by noting that
for each $U_k(1)\in\bar{\mathcal{P}(1)}$ also
$U^\dagger_k(1)=[U_k(1)]^{p-1}\in\bar{\mathcal{P}(1)}$ is contained in
$\bar{\mathcal{P}(1)}$. Consequently,
a Hermitian orthonormal basis is obtained via the
transformation~\cite{LawrencePRA04}  
\begin{eqnarray}\label{eq:hermit}
  H(1)&=&(U(1)-U(1)^\dagger)/\sqrt{2}i\nonumber\\
  \bar H(1)&=&(U(1)+U(1)^\dagger)/\sqrt{2}.
\end{eqnarray}
The operators of kind $H$ have spectrum $\mbox{Im}(\omega^a)$ with
$a\in[0,p-1]$, whereas those of kind $\bar H$ have spectrum 
$\mbox{Re}(\omega^a)$ with $a\in[0,p-1]$. 
Since $[H(1),U(1)]=[\bar H(1),U(1)]=0$, the partitioning structure of the
generalized Pauli basis, and  
hence the corresponding structure of MUB, is preserved by the
transformation~\eqref{eq:hermit}. However, 
since Hermicity is not enforced at the level of the $\lambda$ matrix, 
the Hermitian counterpart of the generalized Pauli basis does not
inherit the tensor product structure, 
\begin{eqnarray}\label{eq:notensor}
  H(n)&=&\bigotimes_{i=1}^nU_i(1))= \big(\bigotimes_{i=1}^nU_i(n)-
  \bigotimes_{i=1}^nU^\dagger_i(n)\big)/i\sqrt{2}\nonumber\\
 &\neq&\bigotimes_{i=1}^nH(U_i)\,.
\end{eqnarray}
If on one hand this implies that the spectrum of the Hermitian operators remains
invariant with respect to the number of qupits on the other 
the operator basis
includes non-local measurements. It is easily seen that, regardless of
the number of particles $n$,  the action of 
a Clifford operation $C$ on   the Hermitian basis  is
$CH(U_k)C^\dagger=H(CU_kC^\dagger)$, since $C$ maps $U_k$ into
$CU_kC^\dagger=\omega^iU_{k'}$ with $i\in[0,p-1]$ 
and $U^\dagger_k$ in $(\omega^{i})^* U^\dagger_{k'}$.

In conclusion, a unitary generalization of the Pauli operators
maintains all 
relevant properties of the standard Pauli basis.
Despite losing Hermicity, it can  be employed to
construct a Hermitian operator  basis which,
however, does not obey a tensor product structure
and hence does not correspond to local measurements.
This sets the stage for efficient characterization
of qupit Clifford operations. 
If one uses the unitary generalized Pauli basis,
despite the fact that the operators are non-Hermitian, actual 
measurements can be carried out utilizing the concept of universal
quantum circuits~\cite{BermejoQuantInf14}:
Any measurement of a generalized (non-Hermitian) Pauli operator can be
implemented by  
applying suitable unitary gates to the system coupled to an
auxiliary qudit and performing a projective measurement on the
auxiliary qudit in the standard basis.
The idea of mapping complex spectra to real measurement results
by an appropriate experimental protocol has first been discussed 
for polarization-path qudits with $p=4$~\cite{PaterekPLA07}.
Alternatively to unitary generalized Pauli measurements, the
Hermitized version of the basis,  
Eq.~\eqref{eq:hermit}, can be adopted.  It includes, 
however, non-local measurements.

\section{Efficient characterization of qudit operations}
\label{sec:qudit_reldist}

\subsection{Modifications of the Monte Carlo approach allowing for
  efficient characterization of qudit operations}
\label{subsec:modifiedMC}
When replacing qubits by qudits, only unitary, maximally partitioning
operator bases such as the generalized Pauli basis and their
Hermitized versions 
allow for efficient characterization both in terms of $\langle
N_{exp}\rangle\propto\mathcal O(1)$ and $\mathcal C_{sampl}\propto
\mathcal O(1)$. 
If a unitary operator basis is chosen, a uniform relevance
distribution can be obtained, yielding efficient sampling, by employing a 
complex generalization
of the standard Monte Carlo
approach~\cite{FlammiaPRL11,daSilvaPRL11,ReichPRL13}.  It  
is presented in  Appendix~\ref{sec:MC_complex}.

For a Hermitized basis, the standard Monte Carlo approach
needs to be  modified at the level of the sampling step. With the
standard sampling procedure, efficient sampling cannot  be achieved
since  the relevance distribution of Clifford unitaries in the
Hermitized basis is no longer uniform due to the loss of the tensor
product structure.
We denote the Hermitized basis by $\mathbb{H}=\{\tilde
H_i\}_{i=1}^{d^2}$ 
where the $\tilde H_i$ comprise both  $H_i$ and its Hermitian 
partner $\bar H_i$. 
For Clifford  operations, the relevance distribution
in the Hermitized basis takes on the values
\begin{equation}\label{eq:hrel}
P^j(i,k)=\{\mbox{Re}^2(\omega^a),
\mbox{Im}^2(\omega^a); a\in[0,p-1]\}\,.
\end{equation}
For each input operator $\tilde{H}_i$ there are two possible 
output operators $\tilde{H}_k$, $\tilde{H}_{\bar{k}}$ leading to
non-vanishing expectation values. The following relation holds
\begin{equation}\label{eq:sum}
P^j(i,k)+P^j(i,\bar k)=1.
\end{equation}
It allows for uniform sampling over pairs $k$, $\bar k$, i.e., 
one draws uniformly at random an index $i\in[1,d^2]$, selecting 
the input operator from the set $\mathbb{H}$. Using a 
generalization  of the Gottesman-Knill
theorem~\cite{BermejoQuantInf14}, 
one can  efficiently compute $C\tilde{H}_iC^\dagger$ where $C$ is 
the Clifford operation that shall be certified.
One thus obtains the indices $k$, $\bar{k}$
corresponding to the measurements with non-vanishing expectation
values and the phase factor $\omega^a$ needed to determine the
corresponding value of the relevance distribution. At this point,
a second sampling step according to Table~\ref{tab:binary} is necessary 
to select a single measurement out of $\tilde{H}_k$ and
$\tilde{H}_{\bar{k}}$. 
\begin{table}[tb]
  \centering
  \begin{tabular}{|c||c|c|}
    \hline 
    & $\tilde{H}_{i}\in\mathbb{H}$ & $\tilde{H}_{i}\in\bar{\mathbb{H}}$\tabularnewline
    \hline \hline 
    $\tilde{H}_{k}\in\mathbb{H}$ & $\text{Re}^{2}\left(\omega^a\right)$
    & $\text{Im}^{2}\left(\omega^a\right)$\tabularnewline 
    \hline 
    $\tilde{H}_{\bar{k}}\in\bar{\mathbb{H}}$ &
    $\text{Im}^{2}\left(\omega^a\right)$ &
    $\text{Re}^{2}\left(\omega^a\right)$\tabularnewline 
    \hline 
  \end{tabular}
  \caption{Relevance distribution for the additional binary sampling
    required for the Hermitized version of the unitary operator basis.
    The symbols $\mathbb{H}$ and $\bar{\mathbb {H}}$ denote,
    respectively, the sets of operators 
    of the kind $H$ and $\bar H$.}
  \label{tab:binary}
\end{table}
Such a two-stage sampling is 
independent of system size. Thus, also for a Hermitized
basis, the sampling complexity is 
$\mathcal{C}_{class}\propto\mathcal{O}(1)$ and the classical computational
resources scale polynomially in $n$. 

\subsection{Hierarchy of operator bases}

Our discussion in Section~\ref{sec:opbases} does not only provide
efficient Monte Carlo protocols for the characterization of qudit
operations, it also allows to classify all operator bases according
to which properties of the standard Pauli basis for qubits they 
retain. The hierarchy is summarized in  Table~\ref{tab:res_clifford}. 
\begin{table*}[tb]
  \centering
  \begin{tabular}{|c|c|c|c|c|}
    \hline
    operator basis & $\langle N_{exp}\rangle$ &
    $\mathcal{C}_{sampl}$& local measurements&protocols \\     
    \hline
    A& $\mathcal{O}(d^2)$& $\mathcal{O}(n^2d^4)$ &yes&1 \\
    \hline
    B &  $\mathcal{O}(1)$ &  as for general unitaries~\footnote{The scaling
      for general unitaries 
      depends on the protocol, cf. Ref.~\cite{ReichPRL13}. }
    & yes&1,2\\ 
    \hline
    C&  $\mathcal{O}(1)$ & as for general unitaries~\footnote{
      If a Hermitian basis comprises non-local measurements, then
      the sampling complexity for general unitaries is increased
      since the relevance distribution can no longer be computed using conditional
      probabilities, cf. Ref. \cite{daSilvaPRL11}. 
    }
   &most likely not&1,2,3\\ 
    \hline
    D&  $\mathcal{O}(1)$ & $\mathcal{O}(1)$ &yes&1,2,3\\
    \hline
    E&  $\mathcal{O}(1)$ & $\mathcal{O}(1)$ &no&1,2,3\\
    \hline
  \end{tabular}
  \caption{Resources required for characterizating of operations
  in $\mathbb{U}_C$. The protocols refer to 1: protocol based on the
  entanglement fidelity~\cite{FlammiaPRL11,daSilvaPRL11},  
  2: protocol employing two classical fidelities~\cite{ReichPRL13},
  3: protocol based on a state 2-design~\cite{ReichPRL13}. The
  operator bases are labeled as follows:
  A: Hermitian bases, such as the Gell-Mann basis for qutrits; 
  B: Hermitian bases constructed as tensor products of 
  a single-qupit bases that give rise to a maximal partitioning with
  all $\lambda^j$ in Eq.~\eqref{eq:specdecomp} being equal;
  C: Hermitian bases that give rise to a maximal partitioning and have 
  equal $\lambda^j$ for all $n$; D: unitary bases that give rise to
  maximal partitioning and have  equal $\lambda^j$ for all $n$, such
  as the generalized Pauli basis; E: Hermitized
  version of D.}  
  \label{tab:res_clifford}
\end{table*}

At the bottom of the hierarchy we find operator bases
that only retain Hermicity, such as the Gell-Mann basis for qutrits.
Following Theorem~1, these bases do not allow for efficient 
Monte Carlo characterization for 
any unitary. Moreover, they cannot be used in combination with the
input-state based protocols that yield a reduction of
resources for general unitaries~\cite{ReichPRL13}. This follows from
the fact that these bases do not allow for the existence of mutually
unbiased eigenbases.

The next step in the hierarchy is occupied by Hermitian bases 
that obey the conditions of Theorem 1.
These bases allow for the existence of a set of non-entangling
unitaries that can be characterized with 
$\langle N_{exp}\rangle\propto\mathcal{O}(1)$ in the protocol based
on the entanglement fidelity and the one using two classical
fidelities. In other words, Theorem 1 ensures that the operator basis
admits the existence of non-entangled generalized stabilizer
states. This explains why the protocol based on a state 2-design 
which includes entangled stabilizer states cannot be applied.
However, the unitaries for which 
$\langle N_{exp}\rangle\propto\mathcal{O}(1)$
cannot be characterized efficiently since in general
their relevance distribution is not known a priori. Therefore, 
Monte Carlo characterization with such operator bases still 
requires classical computational resources that scale exponentially in
the number of qudits.

Next, we have Hermitian operator bases which obey
the conditions of Theorem 2. These bases 
enlarge the class of unitaries for which 
$\langle N_{exp}\rangle\propto\mathcal{O}(1)$ 
to comprise also entangling operations. They also ensure 
that this scaling is achieved in all protocols.
In other words, enforcing the maximally partitioning property
and the condition that all $\lambda$ must be equal
for every $n$ guarantees the existence
of both separable and entangled stabilizer states.
However, most likely, a Hermitian basis for multiple qudits 
which is maximally partitioning includes non-local measurements.
This would imply
that there is no local Hermitian measurement basis allowing
to achieve $\langle N_{exp}\rangle\propto\mathcal{O}(1)$
in all protocols.
Moreover, even if such a basis existed, it would not allow for
efficient characterization  
of any unitary in terms of the sampling complexity
since the relevance distribution  would not be
known a priori. 

Finally, on top of the hierarchy, we find unitary bases that give rise
to a 
maximal partitioning. These bases retain all the relevant properties
of the standard Pauli basis for qubits besides Hermicity. They allow
for efficient characterization in all protocols, provided one
generalizes the Monte Carlo procedure to operators with complex
eigenvalues. The corresponding 
class of unitaries comprises not only the elements of
$\mathbb{U}^0_\delta(n)$, mapping elements of two bases into each
other,  but also certain, if not all, permutations.
Efficient Monte Carlo characterization is also achieved by a
Hermitized version  of such a unitary basis by modifying the sampling
to consist of two stages as explained above.
The Hermitized version, however, comprises
non-local measurements. For generic unitaries, 
Monte Carlo characterization using Hermitized operator bases
requires more computational resources
compared to the unitary counterpart. This is 
due to the loss of the tensor product structure because of which the 
method of the conditional probabilities~\cite{daSilvaPRL11} 
can not be applied.

\section{Conclusions}
\label{sec:concl}

We have shown that there exists a class of unitary operations 
for multi-level information carriers for which in principle the
average fidelity can be estimated efficiently, i.e., with an effort
that scales at most polynomially in the number of qudits. However, if
the class of unitaries is to comprise entangling operations, 
the operator basis that must be chosen to allow for efficient
characterization is either unitary 
non-Hermitian or Hermitian but comprising non-local measurements. 

Unitary non-Hermitian measurements can be realized via quantum
circuits~\cite{PaterekPLA07,BermejoQuantInf14}. 
The corresponding Monte Carlo sampling procedure that is
required to carry out the characterization needs to be adapted to
complex eigenvalues in the relevance distribution. We have shown that
this is straightforward. Employing non-local Hermitian measurements
that are constructed out of the unitary operator basis also requires a small
modification of the standard Monte Carlo procedure in that a two-stage
sampling becomes necessary to achieve a sampling complexity that is
independent of system size. Which of the two approaches, unitary
circuit measurements or non-local Hermitian measurements, is more
practical in an actual experiment remains to be seen. 

The crucial feature of operator bases to allow for efficient
device characterization is that they give rise to a maximal
partitioning of the operators into commuting sets. 
Fulfilling this condition at the level of single-qupit operators
guarantees the existence of a class of unitary transformations that
can be characterized with reduced resources in the Monte Carlo
protocols based on the entanglement
fidelity~\cite{FlammiaPRL11,daSilvaPRL11} and two classical
fidelities~\cite{ReichPRL13}. In that case, a Hermitian basis of local
measurements can be utilized. However, in order to achieve efficient
characterization for a larger set of unitaries including entangling
operations, the maximally partitioning property needs to be fulfilled
at the level of the multi-qudit operators.
While it is automatically satisfied by a  unitary basis built as
tensor product of single-qupit operators that give rise to 
a maximal partitioning, the same does not appear to be true for
Hermitian bases. For the latter, non-local measurements seem
unavoidable for efficient characterization of qudit operations. 

Our work highlights the intimate relation between the
existence of unitaries that can be characterized efficiently and the
existence of mutually unbiased bases. In fact, for prime Hilbert space
dimensions, that is, at the single qupit level, one can determine a
maximal number of such unitaries in a constructive
proof~\cite{Daniel14}. Moreover, 
our results suggest that the conditions presented in Theorems~1 to 3 are
not only sufficient for efficient characterization but also
necessary. One might argue that 
necessity of the maximally partitioning property is questioned by
recent results on generalized Pauli bases~\cite{BermejoQuantInf14}. 
Indeed, a generalized Pauli basis, and 
hence a generalized Clifford group, can be constructed 
assuming only an arbitrary tensor product
decomposition of the Hilbert space, without the necessity of prime
subspace dimensions~\cite{BermejoQuantInf14}. Since existence of a
maximal number of mutually unbiased bases and hence existence of a
maximal partitioning is only guaranteed for prime dimensions, such a
generalized Clifford group would not be in correspondence with an
underlying maximal partitioning structure already at the level of
single qudit operators. 
We believe however that this apparant contradiction can be resolved
by considering the tensor product structure assumed in
Ref.~\cite{BermejoQuantInf14}. Indeed, the properties of a unitary
operator basis that is obtained in terms of tensor products
over an arbitrary decomposition of the Hilbert space should be
equivalent to the properties of the same unitary basis obtained as
tensor products over the irreducible decomposition given by the prime
factorization. 
In the irreducible decomposition, each single-qupit generalized Pauli
basis gives rise to a 
maximal partitioning and thus allows for the existence of stabilizer
states. This would be  consistent with an extension of our theorems in
terms of \textit{necessary} conditions for efficient 
characterization. A rigorous proof of the fact that 
necessity of the maximal partitioning is consistent with the results
of Ref.~\cite{BermejoQuantInf14} is beyond the scope of our current work.

\begin{acknowledgments}
  GG acknowledges financial support from a MIUR-PRIN grant
  (2010LLKJBX). QSTAR is the MPQ, LENS, IIT, 
  UniFi Joint Center for Quantum Science and Technology in Arcetri.
\end{acknowledgments}

\appendix

\section{Proofs}
\label{sec:proofs}

\subsection{Proof of proposition 1}
\label{subsec:proof_prop1}
The general form of a unitary transformation between two bases, $A_j,
A_{j'}\in\mathcal{M}$ is given by Eq.~\eqref{eq:ugen}. This expression is 
general since no ordering of the elements within each basis
is specified. What then needs to be proven is that a change of basis
between $A_j$ and $A_{j'}$ depends only on the distance between the two
indices $j$ and $j'$,  i.e.,
\begin{equation}\label{eq:mainst}
  U_{jj'}=U_\delta\,,
\end{equation}
where $\delta=j-j'$. This can be done by applying $U_{jj'}$ to a generic
element $\ket{\psi^{i}_l}$ of the basis $A_i$ with $i\in[1,d+1]$ and $l\in[1,d]$, 
\begin{equation}\label{eq:appl}
  U_{jj'}\ket{\psi^{i}_l}=\sum_{k=1}^d\ket{\psi^{j'}_k}\langle\psi^{j}_k\ket{\psi^{i}_l}\,.
\end{equation}
Without loss of generality~\footnote{While, to the best of our
knowledge, no strict proof has been reported that all sets of
mutually unbiased bases obey the form of Eq.~\eqref{eq:decomp} or
an equivalent, we  
believe that the proof still works in the general case.},
one can express $\ket{\psi^i_l}$, using the explicit construction of
the mutually unbiased bases for single 
qupits~\cite{WoottersAnnPhys89,BandyoAlgo02}, 
as follows,
\begin{equation}\label{eq:decomp}
  \ket{\psi^i_l}=\frac{1}{\sqrt{d}}
  \sum_{k=1}^d(\omega^l)^{d-k}(\omega^{j-i})^{s_k}\ket{\psi^j_k}\,,
\end{equation}
where $\omega=\exp(2i\pi/p)$ and
$s_k=\sum_{i=k}^di$. Equation~\eqref{eq:decomp}  implies 
\begin{equation}
  \langle\psi^j_k\ket{\psi^i_l}=\frac{1}{\sqrt{d}}(\omega^l)^{d-k}(\omega^{j-i})^{s_k}\,,
\end{equation}
which, substituted into Eq.~\eqref{eq:appl}, leads to
\begin{equation}\label{eq:appl2}
  U_{jj'}\ket{\psi^C_l}=\frac{1}{\sqrt{d}}
  \sum_{k}\ket{\psi^{j'}_k}(\omega^l)^{d-k}(\omega^{j-i})^{s_k}=\ket{\psi^{i+(j'-j)}_l}\,.
\end{equation}
Since this argument holds for any $i\in\mathcal{M}$ and any
$\delta=j'-j$, one can conclude that  
indeed Eq.~\eqref{eq:mainst} holds.
The same is also true for multiple qudits. This can be shown by using, in
Eq.~\eqref{eq:decomp}, 
the general construction of MUB for multiple qupits~\cite{WoottersAnnPhys89}.

Note that, if a precise ordering of the elements within each basis
is chosen, the transformation $U_{\delta}$ can be rewritten as
\begin{equation}\label{eq:udelta2}
U_{\delta}=\sum_k\ket{\psi^{j+\delta}_{\Pi(k)}}\bra{\psi^j_k}\,,
\end{equation}
where $\Pi(k)$  denotes the action of an arbitrary permutation $\Pi$ on 
the $k$th basis index. This yields a decomposition of 
$U_{\delta}$ in terms of a transformation
\begin{equation}\label{eq:uodelta}
  U^0_\delta=\sum_k\ket{\psi^{j+\delta}_k}\bra{\psi^j_k}
\end{equation}
between the $k$th element of basis $j$ and the $k$th element of basis
$j+\delta$ and permutation $\Pi$ of the elements of any of the two
bases,  that is  
\begin{eqnarray}\label{eq:permdecomp}
  \Pi U^0_\delta&=&\sum_{k'k}\ket{\psi^{j+\delta}_{\pi(k')}}
  \bra{\psi^{j+\delta}_{k'}}{\psi^{j+\delta}_k}\rangle\bra{\psi^j_k}=U_\delta\,.
\end{eqnarray}

\subsection{Proof that the unitaries defined by Eq.~\eqref{eq:ugen}
  form a group}
\label{subsec:Ugen_group}

The unitaries defined by Eq.~\eqref{eq:ugen}
form a group, 
$\mathbb{U}^{\Pi}_\delta=\{U_\delta\}_{\delta=0}^{d+1}$,
under matrix multiplication, 
\begin{eqnarray}\label{eq:group}
U_\delta U_{\delta'}&=&\sum_{k,l=1}^d\ket{\psi^{i+\delta}_k}
\bra{\psi^i_k}\psi^{i+\delta'}_l\rangle\bra{\psi^i_l}\\
&=&\sum_{k,l=1}^d\ket{\psi^{i+\delta}_k}\bra{\psi^i_l}\frac{(\omega^l)^{d-k}}
{\sqrt{d}}(\omega^{-\delta'})^{s_k}\\
&=&\sum_l\ket{\psi^{i+(\delta+\delta')}}\bra{\psi^i_l}=U_{\delta+\delta'},
\end{eqnarray}
where we have used that 
\[
\ket{\psi^{i+(\delta+\delta')}}=\frac{1}{\sqrt{d}}\sum_k
\ket{\psi^{i+\delta}_k}(\omega^l)^{d-k}(\omega^{-\delta'})^{s_k}\,,
\]
and 
\begin{equation*}
  U_\delta U^{\dagger}_{\delta}=U_\delta U_{-\delta}=\openone\,.
\end{equation*}
Following the same argument, one can conclude that, for a fixed ordering of the
elements within each basis, the transformations $U^0_\delta$ also form a group 
$\mathbb{U}^0_\delta$ and that the full group $\mathbb{U}^\Pi_\delta$ arises as
the composition of $\mathbb{U}^0_\delta$ with the group of permutations.

\subsection{Proof of Theorem~1}
\label{subsec:theorem1}

Let us apply a unitary $U^0_\delta(1)$, Eq. \eqref{eq:uodelta}, to a generic
element of the single-qupit operator basis, 
\begin{equation}\label{eq:theo1}
  U^0_\delta(1) B^j_i(1)U^{0}_{-\delta}(1)
  =\sum_{k=1}^d\lambda^j_{ik}\ket{\psi^{j+\delta}_k}\bra{\psi^{j+\delta}_k}
  =\tilde{B}_i(1)\,.
\end{equation} 
By definition, $\tilde{B}_i(1)$ belongs to the operator basis
$\mathbb{B}(1)$. It corresponds to the element $B^{j+\delta}_i(1)$, up
to a phase factor $e^{i\phi_i}$, 
if and only if $\lambda^j=e^{i\phi_{j}}\lambda^{j+\delta}$. 
Since this must be true for every $\delta$,
$\lambda^j$ must be equal to $e^{i\phi_j}\lambda$ for each $j\in[1,d+1]$. 
Now since each commuting set contains the identity, i.e., 
the first row of every $\lambda^j$ is  made up of ones,  
$e^{i\phi_j}=1$ and $\lambda^j=\lambda$ for each $j\in[1,d+1]$.
Since the set of unitaries $\mathbb{U}^0_\delta(1)$ forms a group,
the condition on all $\lambda^j$ to be
equal ensures the existence of a  group of transformations which
leaves the single-qupit  operator basis invariant,  i.e., 
$\mathbb{U}^0_\delta(1)\subseteq\mathbb{U}_C(1)$.  

Now consider the $n$-qupit operator basis $\mathbb{B}(n)$, built  
out of tensor products of the operators in $\mathbb{B}(1)$.
The $n$-qupit operators can be written as $B_i(n)=
\bigotimes_{l=1}^nB^{j_l}_{i_l}(1)$, where
$B^{j_l}_{i_l}(1)$ denotes a generic single-qupit operator acting 
on the $l$th qupit. 
Existence of  the group $\mathbb{U}^0_\delta(1)$
implies that $\mathbb{B}(n)$ is left invariant by  the set of unitaries
$\tilde{\mathbb{U}}^0_\delta=\{\tilde{U}^0_{\delta}(n)\}$ that are built
as tensor products of the elements 
in $\mathbb{U}^0_\delta(1)$. This can be seen as follows:
For every $B_i(n)$ and $\tilde{U}^0_{\delta}(n)\in\tilde{\mathbb{U}}^0_\delta$, 
one has
\begin{eqnarray}\label{eq:tens}
\tilde{U}^0_{\delta}(n)&\!B_i(n)\!&\tilde{U}^{0,\dagger}_{\delta}(n)
= \left(\bigotimes_{l=1}^n U^0_{\delta_l}(1)\right)B_i(n)
  \left(\bigotimes_{l=1}^n U^0_{\delta_l}(1)\right)^\dagger \nonumber \\&=&
  \left(\bigotimes_{l=1}^n U^0_{\delta_l}(1)\right)\bigotimes_{l=1}^nB^{j_l}_{i_l}(1)
  \left(\bigotimes_{l=1}^n U^0_{-\delta_l}(1)\right)\nonumber\\
  &=& \bigotimes_{l=1}^n\left(U^0_{\delta_l}(1)B^{j_l}_{i_l}(1)U^0_{-\delta_l}(1)\right)
  \nonumber\\
  &=&  \bigotimes_{l=1}^nB^{j'_l}_{i'_l}(1)=B_{i'}(n)\in\mathbb{B}(n)\,.
 \end{eqnarray}
This allows to conclude
that $\tilde{\mathbb{U}}^0_\delta(n)\subseteq\mathbb{U}_C(n)$.
The characteristic function of  a generic 
$\tilde{U}^0_\delta(n)\in\tilde{\mathbb{U}}^0_\delta(n)$ is given by 
\begin{eqnarray}\label{eq:Uc0}
\chi_{\tilde{U}^0_\delta(n)}(i,k)&=&
\frac{1}{d}\Tr[B_k(n)\tilde{U}^0_\delta(n)B_i(n)\tilde{U}^{0,\dagger}_{\delta(n)}]\nonumber\\ 
&=&\frac{1}{d}\Tr[B_k(n)B_{i'}(n)]=\delta_{ki'}\,.
\end{eqnarray}
Therefore these unitaries will lead to a relevance distribution
with $d^2=\mathcal{N}$ non-zero elements in the protocol
based on the entanglement fidelity, i.e., formally using input
operators~\cite{daSilvaPRL11,FlammiaPRL11}. With 
Eq. \eqref{eq:expscal}, one then finds
$\langle N_{exp}\rangle\propto\mathcal{O}(1)$.
In addition, $\tilde{\mathbb{U}}^0_\delta(n)$
is itself a group since its elements are tensor
products of the elements of $\mathbb{U}^0_\delta(1)$.
 
Let us now  check the scaling of the transformations
in $\tilde{\mathbb{U}}^0_\delta(n)$ for input-state based 
protocols.
By construction, the operator basis $\mathbb{B}(n)$ 
admits the existence of $p+1$
separable 
mutually unbiased joint eigenbases obtained as tensor products
of the elements of the single-qupit bases in $\mathcal{M}(1)$.
These $p+1$ MUB form a subset  $\mathcal{M}_{sep}(n)$ of the full 
set $\mathcal{M}(n)$. By construction, $\mathcal{M}_{sep}(n)$ is mapped into 
itself by the group
of transformations $\tilde{\mathbb{U}}^0_\delta(n)$. Now consider 
a generic element $\ket{\psi^j_i}$ of a separable basis $A_j$ in
$\mathcal{M}_{sep}(n)$. By denoting by $\ket{\psi^{j_l}_{i_l}}$ an
element of the joint eigenbasis of the commuting set
$\mathcal{W}_{j_l}$ of single-qupit operators acting 
on the $l$th qupit,   
$\ket{\psi^j_i}$ can be expressed as 
$\ket{\psi^j_i}=\otimes_{l=1}^n\ket{\psi^{j_l}_{i_l}}$. For each state in $\mathcal{M}^{sep}(n)$, 
the characteristic function
of a unitary transformation $\tilde U^0_\delta(n)\in\tilde{\mathbb{U}}^0_\delta(n)$  is then
\begin{eqnarray}\label{eq:exp}
&&\Tr\left[B_i(n)\tilde U^0_\delta(n)\ket{\psi^j_k}\bra{\psi^j_k}\tilde{U}^{0,\dagger}_\delta(n)\right]\nonumber\\
&=&
 \Tr\left[B_i(n)\ket{\psi^{j'}_k}\bra{\psi^{j'}_k}\right]\nonumber\\ 
 &=& \Pi_{l=1}^n
  \Tr\left[B^{j^{''}_l}_{i_l}(1)\ket{\psi^{j'_l}_{k_l}}\bra{\psi^{j'_l}_{k_l}}\right]\nonumber \\
  &\!\!=\!\!&\begin{cases} 
    E_i(n)&\! \!\mbox{if}\;\; j^{''}_l\!=\!j'_l\;\forall\; l\in[1,n],\\
    0&\mbox{otherwise}
  \end{cases} \,.
\end{eqnarray}
Here $E_i(n)=\Pi_{l=1}^n\lambda^{j'_l}_{i_l,k_l}$ 
is the eigenvalue of $B_i(n)$ corresponding to the element
$\ket{\psi^{j'}_i}$ of the basis $A_{j'}\in\mathcal{M}^{sep}(n)$.
Provided that the characterization protocol does
not require more than $p+1$ MUB, Eq.~\eqref{eq:exp} implies that the
unitaries in
 $\tilde{\mathbb{U}}^0_\delta(n)$ correspond 
 to a relevance distribution with $\mathcal{N}=Td$ non-zero
elements hence yielding 
 $\langle
N_{exp}\rangle\propto\mathcal O(1)$. 
This is the case of  the protocol 
based on classical fidelities since it requires input states from two
MUB but not of the 2-design protocol which instead requires
the existence of the full set $\mathcal{M}(n)$.
 
In conclusion, we have proven that, if the maximally partitioning property and the
condition that all $\lambda^j$'s must be equal are enforced
on the single-qupit operator basis, then the existence for any number of qupit of a non-trivial
group of unitaries leading to $\langle N_{exp}\rangle\propto\mathcal{O}(1)$, at least in some
protocols, is ensured.

\section{Complex Monte Carlo estimation}
\label{sec:MC_complex}

We abbreviate the values of the characteristic functions,
Eq.~\eqref{eq:chi}, by 
\begin{eqnarray*}
\alpha_{ik} &=& \frac{1}{d}\text{Tr}\left[\mathcal{D}\left(W_{i}\right)^{\dagger}W_{k}\right] = \chi_{\mathcal{D}}\left(i,k\right)\\
\beta_{ik} & =&  \frac{1}{d}\text{Tr}\left[U W^{\dagger}_{i} U^{\dagger} W_{k}\right] = \chi_{\mathcal{U}}\left(i,k\right)\,.
\end{eqnarray*}
In general, $\alpha_{ik}$ and $\beta_{ik}$
are complex; they are real only if $W_{k}$ is Hermitian. 
The average gate fidelity can be expressed in terms of $\alpha_{ik}$
and $\beta_{ik}$, 
\begin{eqnarray*}
F_{av}&=&\frac{1}{d^{2}}\sum_{i,k}\alpha_{ik}\beta_{ik}^{*}
=\sum_{i,k}\frac{\left|\beta_{ik}\right|^{2}}{d^{2}}\frac{\alpha_{ik}}{\beta_{ik}}\\
&=&\sum_{i,k}\text{Pr}\left(i,k\right)\frac{\alpha_{ik}}{\beta_{ik}}  
\end{eqnarray*}
with the real-valued relevance distribution
\[
\text{Pr}\left(i,k\right)=\frac{\left|\beta_{ik}\right|^{2}}{d^{2}}\,.
\]
Note that if $U_{0}$ is a Clifford gate, then for any $i$ there is only
a single $k$ such that $\beta_{ik}\neq0$, taking the value
$\frac{1}{d^{2}}$.
For Monte Carlo sampling we define now the complex random variable
$X$ on the event space given by the set of tupels $\left(i,k\right)$
\begin{equation}
X\left(i,k\right)=\frac{\alpha_{ik}}{\beta_{ik}}\,.
\end{equation}
It is easy to see that the expectation value of this random variable
corresponds to $F_{av}$, 
\begin{equation}
\mathbb{E}\left(X\left(i,k\right)\right)
=\sum_{i,k}\text{Pr}\left(i,k\right)\frac{\alpha_{ik}}{\beta_{ik}}=F_{av}\,. 
\end{equation}

The Monte Carlo approach seeks an estimate of $F_{av}$ 
with additive error $\epsilon$ and failure probability 
$\delta$. In other words, one wants to find an estimator $Y$ such that
the likelihood that this estimator $Y$ is greater or equal $\epsilon$
away from the fidelity $F_{av}$ to be less or equal $\delta$,
\begin{equation}
  \text{Pr}\left[\left|Y-F_{av}\right|\geq\epsilon\right]\leq\delta\label{eq:cheby}\,.
\end{equation}
The complex version of Chebyshev's inequality~\cite{Cheby} states
that, $\forall t>0$ and each complex random 
variable $Z$ with expectation value $\mu$, the following relation
is fulfilled 
\begin{equation}
  \text{Pr}\left[\left|Z-\mu\right|
    \geq t\left|\mu\right|\right]
  \leq\frac{\mathbb{E}\left(ZZ^{*}\right)
    -\mathbb{E}\left(Z\right)\mathbb{E}\left(Z^{*}\right)}
  {t^{2}\left|\mu\right|^{2}}\label{eq:cheby2}\,.
\end{equation}
Mapping $t>0$ onto $t\left|\mu\right|\equiv\kappa>0$ leads to
\begin{equation}
  \text{Pr}\left[\left|Z-\mu\right|
    \geq\kappa\right]
  \leq\frac{\mathbb{E}\left(ZZ^{*}\right)
    -\mathbb{E}\left(Z\right)\mathbb{E}\left(Z^{*}\right)}
  {\kappa^{2}}\label{eq:cheby3}\,.
\end{equation}
Now one just needs to find a suitable estimator $Y$ and calculate
its expectation value and variance.

To this end, set the number of draws $L$ from the event space
given by the tupels $\left(i,k\right)$ to $L=\left\lceil
  \frac{1}{\epsilon^{2}\delta}\right\rceil $ where 
$\left\lceil \cdot\right\rceil $ means to round up to the nearest
integer. Choosing independently 
some events $\left(i_{1},k_{1}\right),\dots,\left(i_{L},k_{L}\right)$
out of the total event space 
yields independent estimates
$X_{1}=\frac{\alpha_{i_{1}k_{1}}}{\beta_{i_{1}k_{1}}},
\dots,X_{L}=\frac{\alpha_{i_{L}k_{L}}}{\beta_{i_{L}k_{L}}}$. 
Now define $Y=\frac{1}{L}\sum_{l=1}^{L}X_{l}$. We explain
how to estimate $Y$ which in turn is an approximation to $F_{av}$. Note
that $Y$ structurally resembles $X$. However, the relevance
distribution does not appear. This is due to the fact that
each $X_{l}$ is already
chosen with the corresponding probability.
Hence in the limit of $L\rightarrow\infty$:
$Y\rightarrow X$.

Consider the choice of $\left(i_{l},k_{l}\right)$ with $l=1,\dots,L$
chosen as explained above and $i_{l}$ denoting 
the index of the input operator of the $l$th measurement
by $k_l$ the index of the measured operator of the $l$th measurement.
For each $l$ the operator $W_{k_{l}}$ will be measured on the state
that is obtained by sending a randomly drawn eigenstate
$\ket{\phi_{a}^{i_{l}}}$ of $W_{i_{l}}$ with corresponding eigenvalue
$\lambda_{a}^{i_{l}}$ through the device  
($a$ is drawn out of the set $\left\{ 1,\dots,d\right\} $). This
is repeated a total number of $m_{l}$ times where 
\begin{equation}
m_{l}=\left\lceil \frac{4}{\left|\beta_{i_{l}k_{l}}\right|^{2}L\epsilon^{2}}\log\left(\frac{4}{\delta}\right)\right\rceil\, .\label{eq:number of experiments}
\end{equation}
This choice of $m_{l}$ guarantees that Eq.~\eqref{eq:cheby} is
fulfilled as 
we show below. Note that each measurement gives an eigenvalue
of the operator $W_{k_{l}}$. We denote these, in general complex, 
measurement results by $w_{ln}$ with $n$ referring to 
the $n$th repetition of the $l$th measurement. Each
of these measurements results in an eigenvalue
$w_{ln}\in\text{spec}\left(W_{k}\right)$. 
We assume the expectation value of
a measurement of an operator $W_{k_{l}}$ for a state 
$\rho$ to be given by 
\[
\left\langle W_{k_{l}}\right\rangle _{\rho}=\text{Tr}\left[\rho^{\dagger}W_{k_{l}}\right]=\text{Tr}\left[\rho W_{k_{l}}\right]
\]
also for non-Hermitian operators.
Let us define now
$A_{ln}=\left(\lambda_{a_{n}}^{i_{l}}\right)^{*}w_{ln}$
where $\lambda_{a_{n}}^{i_{l}}$ is the eigenvalue corresponding to
the eigenstate $\ket{\phi_{a_{n}}^{i_{l}}}$ of the operator $W_{i_{l}}$.
Note that
\begin{eqnarray*}
  \mathbb{E}\left(A_{ln}\right) & = &
  \frac{1}{d}\sum_{a_{n}=1}^{d}\left(\lambda_{a_{n}}^{i_{l}}\right)^{*}w_{ln}\\
  &=&\frac{1}{d}\sum_{a_{n}=1}^{d}\left(\lambda_{a_{n}}^{i_{l}}\right)^{*}\text{Tr}\left[\mathcal{D}\left(\ket{\phi_{a_{n}}^{i_{l}}}\bra{\phi_{a_{n}}^{i_{l}}}\right)^{\dagger}W_{k_{l}}\right]\\
&=&\frac{1}{d}\sum_{a_{n}=1}^{d}\text{Tr}\left[\left(\lambda_{a_{n}}^{i_{l}}\right)^{*}\mathcal{D}\left(\ket{\phi_{a_{n}}^{i_{l}}}\bra{\phi_{a_{n}}^{i_{l}}}\right)^{\dagger}W_{k_{l}}\right]\\
 &
 =&\frac{1}{d}\text{Tr}\left[\mathcal{D}\left(\sum_{a_{n}=1}^{d}\lambda_{a_{n}}^{i_{l}}\ket{\phi_{a_{n}}^{i_{l}}}\bra{\phi_{a_{n}}^{i_{l}}}\right)^{\dagger}W_{k_{l}}\right]\\
&=&\frac{1}{d}\text{Tr}\left[\mathcal{D}\left(W_{i_{l}}\right)^{\dagger}W_{k_{l}}\right]=\alpha_{i_{l}k_{l}}\,.
\end{eqnarray*}
An approximation to $X_{l}$, denoted by  $\tilde{X}_{l}$, can now be
introduced, 
\begin{equation}
\tilde{X}_{l}=\frac{1}{\beta_{i_{l}k_{l}}}\cdot\frac{1}{m_{l}}\sum_{n=1}^{m_{l}}A_{ln}\,.
\label{eq:Xl}
\end{equation}
Since  $\mathbb{E}\left(B_{ln}\right)\equiv\left\langle
  A_{ln}\right\rangle =\alpha_{i_{l}k_{l}}$, 
it is clear that $\frac{1}{m_{l}}\sum_{n=1}^{m_{l}}A_{ln}\rightarrow\alpha_{i_{l}k_{l}}$.

For the final step in the Monte Carlo estimation, let 
\begin{equation}
\tilde{Y}=\frac{1}{L}\sum_{l=1}^{L}\tilde{X}_{l}\,.
\end{equation}
Just like  $\tilde{X}_{l}$ is an approximation
to $X_{l}$,  $\tilde{Y}$ is an approximation
to $Y$ or in other words an estimate for $Y$. The goal is to fulfill
Hoeffding's inequality, which we prove below,
\begin{equation}
\text{Pr}\left[\left|\tilde{Y}-Y\right|\geq\epsilon\right]\leq\delta\label{eq:hoeffding}\,. 
\end{equation}
The whole procedure uses the channel a total number of $m=\sum_{l=1}^{L}m_{l}$
times. This value in estimation can be bounded by calculating
$\mathbb{E}\left(m_{l}\right)$ 
which is the expected number of experimental repetitions for the given
setting $\left(i_{l},k_{l}\right)$. In other words $\mathbb{E}\left(m_{l}\right)$
is the number of experiments one has to perform for a setting $\left(i,k\right)$
multiplied by the probability that this setting is chosen.
Denoting by $m_{l}\left(i,k\right)$
the number of experiments for the tupel $\left(i,k\right)$, given
by Eq.~\eqref{eq:number of experiments}, the expectation value becomes  
\begin{eqnarray}
  \mathbb{E}\left(m_{l}\right) & = &
  \sum_{ik}\text{Pr}\left(i,k\right)m_{l}\left(i,k\right)\nonumber \\
  &=&\frac{1}{d^{2}}\sum_{ik}\left|\beta_{ik}\right|^{2}\left\lceil \frac{4}{\left|\beta_{ik}\right|^{2}L\epsilon^{2}}\log\left(\frac{4}{\delta}\right)\right\rceil \label{eq:experiments number}\\
 & \leq & 1+\frac{4d^{2}}{L\epsilon^{2}}\log\left(\frac{4}{\delta}\right)\nonumber\,,
\end{eqnarray}
where $1$ accounts for the fact that the smallest
integer greater than the expression in the brackets $\left\lceil
  \cdot\right\rceil $ is taken.
The total number of experiments given by the sum of all $m_{l}$ is found to be 
\begin{eqnarray}
  \mathbb{E}\left(m\right)&=&
  \sum_{l=1}^{L}\mathbb{E}\left(m_{l}\right)\leq
  L\cdot\left[1+\frac{4d^{2}}{L\epsilon^{2}}\log\left(\frac{4}{\delta}\right)\right]\nonumber\\
&\leq&1+\frac{1}{\epsilon^{2}\delta}+\frac{4d^{2}}{\epsilon^{2}}\log\left(\frac{4}{\delta}\right)  \,,
\end{eqnarray}
where $1$ appears for the same reason as above. Note
that this scales as $\mathcal{O}\left(d^{2}\right)$. For 
Clifford gates, there are only $d^{2}$ nonvanishing entries
the sum in Eq.~\eqref{eq:experiments number} since for 
each $k$ there exists only one $l$ for which $\beta_{kl}\neq0$. This
leads to 
\[
\mathbb{E}\left(m_{l}\right)\leq1+\frac{4}{L\epsilon^{2}}\log\left(\frac{4}{\delta}\right)
\]
and consequently
\[
\mathbb{E}\left(m\right)\leq1+\frac{1}{\epsilon^{2}\delta}+\frac{4}{\epsilon^{2}}\log\left(\frac{4}{\delta}\right)\,,
\]
resulting in a scaling of $\mathcal{O}\left(1\right)$.

Finally we prove validity of Eqs.~\eqref{eq:cheby} and~\eqref{eq:hoeffding}.
We first consider Eq.~\eqref{eq:cheby}, where the numerator of the
right hand side of the Chebyshev inequality needs to be estimated
for $Z=X_{l}$, 
\begin{widetext}
\begin{eqnarray*}
\mathbb{E}\left(X_{l}X_{l}^{*}\right)-\mathbb{E}\left(X_{l}\right)\mathbb{E}\left(X_{l}^{*}\right) & = & \sum_{ik}\text{Pr}\left(i,k\right)\frac{\left|\alpha_{ik}\right|^{2}}{\left|\beta_{ik}\right|^{2}}-\left|\sum_{ik}\text{Pr}\left(i,k\right)\frac{\alpha_{ik}}{\beta_{ik}}\right|^{2}
 =  \frac{1}{d^{2}}\sum_{ik}\left|\alpha_{ik}\right|^{2}-F^{2}\\
 & = & \frac{1}{d^{4}}\sum_{ik}\langle\langle\mathcal{D}\left(W_{i}\right)\|W_{k}\rangle\rangle\langle\langle W_{k}\|\mathcal{D}\left(W_{i}\right)\rangle\rangle-F_{H}^{2}
 =  \frac{1}{d^{4}}\sum_{ik}\left|\text{Tr}\left[W_{k}^{\dagger}\mathcal{D}\left(W_{i}\right)\right]\right|^{2}-F^{2}
\end{eqnarray*}
\end{widetext}
Obviously $0\leq F\leq1\Longrightarrow0\leq F^{2}\leq1$ for all fidelities discussed in this paper.
The same is true for the first term. This can be seen most
easily in terms of the process matrix. For any operator $O$, one can
write 
\[
\mathcal{D}\left(O\right)=\sum_{nm}\chi_{nm}W_{m}OW_{n}^{\dagger}\,.
\]
Clearly  for $O=W_{i}$, 
\[
\mathcal{D}\left(W_{i}\right)=\sum_{nm}\chi_{nm}W_{m}W_{i}W_{n}^{\dagger}\,.
\]
It follows that
\begin{eqnarray*}
\left|\text{Tr}\left[W_{k}^{\dagger}\mathcal{D}\left(W_{i}\right)\right]\right|^{2}
&=&\left|\sum_{nm}\chi_{nm}\text{Tr}\left[W_{k}^{\dagger}W_{m}W_{i}W_{n}^{\dagger}\right]\right|^{2}\\
&\leq&\sum_{nm}\left|\chi_{nm}\right|^{2}\left|\text{Tr}\left[W_{k}^{\dagger}W_{m}W_{i}W_{n}^{\dagger}\right]\right|^{2}  
\end{eqnarray*}
For fixed $i$ and $k$, the operator $W_{k}^{\dagger}W_{m}W_{i}$
is proportional to a Pauli operator. Consider the expression
\[
\sum_{ik}\left|\text{Tr}\left[W_{k}^{\dagger}W_{m}W_{i}W_{n}^{\dagger}\right]\right|^{2}\,.
\]
For fixed $m,n$ and a certain $i$ there exists exactly one $k$
such that this is nonzero, namely if and only if
\begin{equation}
W_{k}^{\dagger}W_{m}W_{i}W_{n}^{\dagger}\sim\openone_{d}\label{eq:proportionality}\,.
\end{equation}
That is,
\[
W_{k}\sim W_{m}W_{i}W_{n}^{\dagger}\,.
\]
Due to orthonormality of the operator basis, there is only
one such $k$ for which this relation can be fulfilled. For Pauli operators
the proportionality constant has modulus $1$, hence
\[
\sum_{ik}\left|\text{Tr}\left[W_{k}^{\dagger}W_{m}W_{i}W_{n}^{\dagger}\right]\right|^{2}
=d^{2}\cdot d^{2}=d^{4}\,.
\]
This results in a trace of $d$ for the $d^{2}$ tupel $\left(i,k\right)$
for which relation \eqref{eq:proportionality} holds. Consequently
\[
\frac{1}{d^{4}}\sum_{ik}\left|\text{Tr}\left[W_{k}^{\dagger}\mathcal{D}\left(W_{i}\right)\right]\right|^{2}\leq\sum_{ik}\left|\chi_{ik}\right|^{2}\,.
\]
Due the Choi-Jamiolkowsky isomorphism, the process
matrix corresponds to a density matrix in the $d^{2}$-dimensional
Hilbert space $\mathcal{H}\otimes\mathcal{H}$. It can easily be seen
that $\sum_{ik}\left|\chi_{ik}\right|^{2}$ corresponds to the purity
of this density matrix which cannot be greater than $1$. Therefore 
\[
\frac{1}{d^{4}}\sum_{ik}\left|\text{Tr}\left[W_{k}^{\dagger}\mathcal{D}\left(W_{i}\right)\right]\right|^{2}\leq1\,.
\]
Hence $\left[\mathbb{E}\left(X_{l}X_{l}^{*}\right)-\mathbb{E}\left(X_{l}\right)\mathbb{E}\left(X_{l}^{*}\right)\right]$
is the difference between two numbers in the interval $\left[0,1\right]$
and consequently smaller than $1$, 
\[
\mathbb{E}\left(X_{l}X_{l}^{*}\right)-\mathbb{E}\left(X_{l}\right)\mathbb{E}\left(X_{l}^{*}\right)\leq1\,.
\]
It follows for $Y=Y=\frac{1}{L}\sum_{l=1}^{L}X_{l}$ that
\begin{widetext}
\begin{align*}
\mathbb{E}\left(YY^{*}\right)-\mathbb{E}\left(Y\right)\mathbb{E}\left(Y^{*}\right) & =\mathbb{E}\left(\left(\frac{1}{L}\sum_{l}X_{l}\right)\left(\frac{1}{L}\sum_{l'}X_{l'}^{*}\right)\right)-\mathbb{E}\left(\frac{1}{L}\sum_{l}X_{l}\right)\mathbb{E}\left(\frac{1}{L}\sum_{l}X_{l}^{*}\right)\\
 & =\frac{1}{L^{2}}\sum_{ll'}\mathbb{E}\left(X_{l}X_{l'}^{*}\right)-\frac{1}{L^{2}}\sum_{ll'}\mathbb{E}\left(X_{l}\right)\mathbb{E}\left(X_{l'}^{*}\right)\\
 & =\frac{1}{L^{2}}\sum_{ll'}\left[\mathbb{E}\left(X_{l}X_{l'}\right)-\mathbb{E}\left(X_{l}\right)\mathbb{E}\left(X_{l'}^{*}\right)\right]
 =\frac{1}{L^{2}}\sum_{l}\left[\mathbb{E}\left(X_{l}X_{l}\right)-\mathbb{E}\left(X_{l}\right)\mathbb{E}\left(X_{l}^{*}\right)\right]\\
 & \leq\frac{L}{L^{2}}=\frac{1}{L}
\end{align*}  
\end{widetext}
where use has been made of $\mathbb{E}\left(X_{l}X_{l'}\right)=
\mathbb{E}\left(X_{l}\right)\mathbb{E}\left(X_{l'}\right)$ for the
$X_{l}\neq X_{l'}$ which are uncorrelated.
Chebyshev's inequality, Eq. \eqref{eq:cheby2},  consequently yields
\begin{equation}
\text{Pr}\left[\left|Y-F\right|\geq\kappa\right]\leq\frac{1}{L\kappa^{2}}\,.
\end{equation}
Now set $\kappa=\sqrt{\frac{1}{L\delta}}$ and $L=\frac{1}{\epsilon^{2}\delta}$
to obtain
\[
\text{Pr}\left[\left|Y-F\right|\geq\epsilon\right]\leq\delta
\]

To show the validity of Eq. \eqref{eq:hoeffding} we use the complex
version of Hoeffding's inequality \cite{Hoeffding}. 

\textbf{Lemma:}
Let $\vec{a}\in\mathbb{R}^{n}$ and $\left\{ X_{i}\right\} _{i=1,\dots,N}$
be independent zero-mean complex-valued random variables with $\forall
i:\ \left|X_{i}\right|\leq a_{i}$. Then $\forall\delta>0$
\[
\text{Pr}\left(\left|\sum_{i=1}^{N}X_{i}\right|\geq\delta\right)\leq4\exp\left(-\frac{\delta^{2}}{4\sum_{i=1}^{n}\left|a_{i}\right|^{2}}\right)
\]

\textbf{Corollary:} Let $\vec{a}\in\mathbb{R}^{n}$ and $\left\{
  X_{i}\right\} _{i=1,\dots,N}$ 
be independent complex-valued random variables with mean value $\sum_{i=1}^{N}\left\langle X_{i}\right\rangle =\left\langle X\right\rangle $
where $X=\sum_{i=1}^{N}X_{i}$ and $\forall i:\ \left|X_{i}-\left\langle
    X_{i}\right\rangle \right|\leq a_{i}$. Then $\forall\delta>0$
\[
\text{Pr}\left(\left|X-\left\langle X\right\rangle \right|\geq\delta\right)\leq4\exp\left(-\frac{\delta^{2}}{4\sum_{i=1}^{n}\left|a_{i}\right|^{2}}\right)
\]

Proof: Apply Hoeffding's inequality to the random variables
$X_{i}-\left\langle X_{i}\right\rangle $.

Specifically this means for $\delta>0$, $n=L$ and
$\tilde{Y}=\frac{1}{L}\sum_{l=1}^{L}\tilde{X}_{l}$ 
with $\left\langle \tilde{Y}\right\rangle =\frac{1}{L}\sum_{l=1}^{L}\left\langle \tilde{X}_{l}\right\rangle =\frac{1}{L}\sum_{l=1}^{L}X_{l}=Y$.
Note furthermore that the $\tilde{X}_{l}$ are composed as a sum themselves
of independent random variables $A_{ln}$ corresponding to measurement
results with modulus smaller than $1$ and expectation value with
modulus smaller than $1$. As such we can write 
\begin{equation}
\text{Pr}\left[\left|\tilde{Y}-Y\right|\geq\epsilon\right]\leq4\exp\left(-\frac{4\epsilon^{2}}{C}\right)
\end{equation}
 where 
\begin{equation}
C=\sum_{l=1}^{L}\frac{1}{L}m_{l}\left|2c_{l}\right|^{2},\quad c_{l}=\frac{1}{m_{l}\beta_{i_{l}k_{l}}}
\end{equation}
since $\left[A_{ln}-\left\langle A_{ln}\right\rangle \right]$, 
as discussed for Eq. \eqref{eq:Xl}, always takes values with modulus
smaller than $2$.

Calculating $C$ leads to 
\begin{eqnarray}
  C&=&\sum_{l=1}^{L}\frac{4}{L^{2}\beta_{i_{l}k_{l}}^{2}m_{l}}
  =\sum_{l=1}^{L}\frac{4\beta_{i_{l}k_{l}}^{2}L\epsilon^{2}}
  {4L^{2}\beta_{i_{l}k_{l}}^{2}\log\left(\frac{4}{\delta}\right)}\nonumber
  \\
  &=&\sum_{l=1}^{L}\frac{\epsilon^{2}}{L\log\left(\frac{4}{\delta}\right)}=\frac{\epsilon^{2}}{\log\left(\frac{4}{\delta}\right)}  \,.
\end{eqnarray}
Plugging this into Hoeffding's inequality yields 
\begin{eqnarray}
\text{Pr}\left[\left|\tilde{Y}-Y\right|\geq\epsilon\right] & \leq & 4\exp\left(-\frac{4\epsilon^{2}}{C}\right)=4\exp\left(-4\log\left(\frac{4}{\delta}\right)\right)\nonumber \\
 & \leq & 4\exp\left(\log\left(\frac{\delta^{2}}{16}\right)\right)=\frac{\delta^{2}}{4}\leq\delta\,.
\end{eqnarray}
Hence the failure probability is $\leq\delta$ as desired.


\end{document}